\def\Msun{\hbox{M$_\odot$}}
\def\Ne{\hbox{$n_{\rm {e}}$}}
\def\Te{\hbox{$T_{\rm {e}}$}}
\def\kms{\hbox{km$\,$s$^{-1}$}}
\def\cmt{\hbox{cm$^{-3}$}}

\def\one{\,{\sc i}}             % for producing Na I as Na\one\ etc.
\def\two{\,{\sc ii}}
\def\three{\,{\sc iii}}

\documentclass{emulateapj}

\usepackage{amsmath}
\usepackage{amssymb}
\usepackage{mathrsfs}
\usepackage{color}
\usepackage{upgreek}
\usepackage{graphicx}

\shorttitle{The environment in the starburst core of M82}
\shortauthors{M.\ S.\ Westmoquette et al.}

\begin{document}
\defcitealias{smith06}{Paper I}
\defcitealias{ab99}{AB99}
\defcitealias{ba99}{BA99}

%%%%%%%%%%%%%%%%%%%%%%%%%%%%%%%%%%%%%%
%AUTHORS
%%%%%%%%%%%%%%%%%%%%%%%%%%%%%%%%%%%%%%

\title{\textit{HST}/STIS spectroscopy of the environment in the starburst core of M82\altaffilmark{1}}
\author{M.\ S.\ Westmoquette\altaffilmark{2}\email{msw@star.ucl.ac.uk}}
\author{L.\ J.\ Smith\altaffilmark{2,3}}
\author{J.\ S.\ Gallagher III\altaffilmark{4}}
\author{R.\ W.\ O'Connell\altaffilmark{5}}
\author{D.\ J.\ Rosario\altaffilmark{5}}
\author{R.\ de Grijs\altaffilmark{6,7}}

\altaffiltext{1}{Based on observations with the NASA/ESA {\it Hubble Space Telescope}, obtained at the Space Telescope Science Institute, which is operated by AURA, Inc., under NASA contract NAS5--26555. These observations are associated with program \#9117.}
\altaffiltext{2}{Department of Physics and Astronomy, University College London, Gower Street, London, WC1E 6BT, UK}
\altaffiltext{3}{Space Telescope Science Institute and European Space Agency, 3700 San Martin Drive, Baltimore, MD 21218, USA}
\altaffiltext{4}{Department of Astronomy, University of Wisconsin-Madison, 5534 Sterling, 475 North Charter St., Madison WI 53706, USA}
\altaffiltext{5}{Department of Astronomy, University of Virginia, P.O. Box 3818, Charlottesville, VA 22903, USA}
\altaffiltext{6}{Department of  Physics and Astronomy, The University of Sheffield, Hicks Building, Hounsfield Rd., Sheffield, \\S3 7RH, UK}
\altaffiltext{7}{National Astronomical Observatories, Chinese Academy of Sciences, 20A Datun Road, Chaoyang District, Beijing 100012, P.R. China}
%

%%%%%%%%%%%%%%%%%%%%%%%%%%%%%%%%%%%%%%
%%%%%%%%%%%%%%%%%%%%%%%%%%%%%%%%%%%%%%
\begin{abstract}
We present high-resolution optical {\it HST} Space Telescope Imaging Spectrograph (STIS) observations made with two slits crossing four of the optically brightest starburst clumps in the vicinity of the nucleus of M82. These provide H$\alpha$ kinematics, extinction, electron density and emission measures. From the radial velocity curves derived from both slits we confirm the presence of a stellar bar. We also find that the super star cluster M82-A1 has a position and radial velocity consistent with it being at the end of one of the unique $x_{2}$ bar orbits formed by an inner Lindblad resonance. We derive a new model for the orientation of the bar and disc with respect to the main starburst clumps, and propose that clump A has formed within the bar region as a result of gas interactions between the bar orbits, whereas region C lies at the edge of the bar and regions D and E are located further out from the nucleus but heavily obscured. We derive extremely high interstellar densities of 500--900~\cmt, corresponding to ISM pressures of $P/k \approx 0.5$--$1.0\times 10^{7}$~\cmt~K, and discuss the implications of the measured gas properties surrounding the nuclear star clusters on the production and evolution of the galactic wind. Despite varying pressures, the ionization parameter is uniform down to parsec-scales, and we discuss why this might be so. Where the signal-to-noise (S/N) of our spectra are high enough, we identify multiple emission-line components. Through detailed Gaussian line-fitting, we identify a ubiquitous broad (200--300~\kms) underlying component to the bright H$\alpha$ line, and discuss the physical mechanism(s) that could be responsible for such widths. We conclude that the evaporation and/or ablation of material from interstellar gas clouds caused by the impact of the high-energy photons and fast-flowing cluster winds produces a highly turbulent layer on the surface of the clouds from which the emission arises.
\end{abstract}

\keywords{galaxies: individual(M82) -- galaxies: ISM -- galaxies: starburst -- galaxies: star clusters -- galaxies: kinematics and dynamics -- ISM: evolution}

%%%%%%%%%%%%%%%%%%%%%%%%%%%%%%%
\section{Introduction}\label{sect:intro}

Galaxies undergoing starburst events, triggered by interactions or mergers, are important objects to study in the local universe because of the insights they provide into the violent star formation processes that occurred in galaxies at much earlier epochs. M82 is the best-studied nearby \citep[3.6~Mpc;][]{freedman94} example of such a galaxy. The current starburst is thought to have been triggered by a tidal interaction with M81 a few $\times$ $10^8$~yrs ago \citep*{yun_ho_lo94}, but investigations of this starburst are hampered by the galaxy's almost edge-on inclination \citep[$i \sim 80^{\circ}$;][]{lynds63,mckeith95}. Much of the starburst core therefore suffers from heavy extinction along the line-of-sight, so most observations have concentrated on wavelengths beyond the visible domain \citep[e.g.][]{rieke80, achtermann95, satyapal95, satyapal97, forster01}.

%SSCs
\citet{oconnell78} identified a number of high surface brightness clumps or regions in M82, denoted A, C, D and E that define the optical starburst core, and cover an area of $\sim$500~pc in diameter. However, the study of starbursts at resolutions attainable with the {\it Hubble Space Telescope (HST)} has revealed the existence of hyper-luminous compact star clusters, so-called super star clusters (SSCs), within sites of intense star-formation. With {\it HST} Planetary Camera imaging, \citet{oconnell95} identified over one hundred candidate SSCs within the visible starburst, while more recently \citet{melo05} catalogued a total of 197 young massive clusters in the starburst core with {\it HST}/WFPC2 observations, and they associated 86 with region A alone. This incredible density of star clusters gives rise to a very unusual, highly energetic environment, and is very different to anything we find in our own Galaxy or even in other less intense starbursts.

%starburst conditions - forster01 + paper I findings
In a companion paper \citep[][hereafter \citetalias{smith06}]{smith06}, we describe observations of an isolated SSC in region A, designated M82-A1, and find it to be surrounded by a surprisingly compact H\two\ region ($r = 4.5\pm 0.5$~pc compared to the cluster radius, $r = 3.5\pm 0.5$~pc) at an unusually high pressure ($P/k=1$--$2\times 10^{7}$~cm$^{-3}$~K). The cluster H\two\ region was found to have an  ionization parameter of $\log U = -2.24\pm 0.18$, in excellent agreement with the IR study of \citet{forster01}. These authors found, despite the seemingly chaotic and rapidly varying nature of the ISM, a uniform $\log U \approx -2.3$ for scales ranging all the way from a few tens to 500~pc, leading them to suggest that a similar star-formation efficiency and evolutionary stage exists across the whole starburst region. Our observation of a similar ionization parameter extends the uniformity of the starburst conditions down to scales of only a few parsecs. An alternative model presented by \citet{dopita02} associates a constant observed $U$ with situations where dust effectively competes with gas in absorbing ionizing photons. We briefly discuss this issue in Section~\ref{sect:emission_meas}. In \citetalias{smith06}, we discussed how such a compact, high density H\two{} region should theoretically not exist at the estimated age of the cluster ($6.4\pm 0.5$~Myr) if it had evolved according to the standard model for a pressure-driven bubble. We found we could explain its existence through the high ambient pressures resulting in non-standard evolution.

%compact SNR
High ambient interstellar pressures have also been inferred from measurements of the size of supernova remnants (SNR) in the core of M82. Radio studies have shown that they are unusually compact \citep[radii $<$4~pc;][]{muxlow94, pedlar99}. This suggests that, together with the results mentioned above, the densities and pressures are the dominant drivers of the H\two{} region evolution and that pressure in the starburst core is a major factor influencing the starburst evolution \citep[see also][]{rigby04}.

%bar
M82 hosts a stellar bar that is thought to have formed during its last encounter with M81 \citep{telesco91, wills00, greve02a}. \citet{wills00} developed a model of the M82 bar system by comparing the predictions of their analytical model to neutral (H\one, CO) and ionized ([Ne\two]12.8\,$\mu$m) gas observations. They found that the ionized gas is dominated by a faster rotating component that they identify with so-called $x_{2}$ bar orbit family \citep[thus corresponding to the ionized ring identified by][]{achtermann95}, whilst the neutral and molecular gas appears more consistent with the slower rotating cusped $x_{1}$ orbit family \citep[$x_{2}$-orbits are associated with an inner Lindblad resonance;][]{athanassoula92a}. \citet{ab99} perform extensive hydrodynamical simulations of bars, concentrating specifically on the gaseous component, and find the gas in fact follows stream-lines that correspond only qualitatively to the $x_{1}$- and $x_{2}$-orbits. Detailed inspection shows that there are differences arising from the fact that in reality the gas stream-lines only loosely resemble the stellar orbit families. In \citet{greve02a}, K.\ Wills and her collaborators used the near-IR Ca\two\,$\lambda8542$ stellar photospheric absorption line observations from \citet{mckeith93} to compare their bar model directly to stellar velocities rather than the previously used gas measurements. By making this comparison, \citeauthor{greve02a}~were able to estimate that the mass of the stars following $x_{2}$-orbits is $\sim$15 per cent of the total mass of the bar.

%ISM model
The densely packed star clusters, high densities and the presence of a bar give rise to unusual, highly energetic ISM conditions. \citet{lord96} developed a model for the ISM in the central 700~pc of M82 based on far-IR spectroscopic observations of forbidden-lines \citep[see also][]{forster01, forster03}. They proposed that the observed emission can be explained as originating in multiple H\two{} regions and PDRs (photo-dissociation regions) mixed uniformly with the ionizing stars and clusters. They found that the clouds have a characteristic size, $r_{\rm cl} \sim 0.4$--1~pc (with a molecular core and extended ionized envelope), density, \Ne{}~$\sim$~$10^{4}$~\cmt{}, and mass range, $M_{\rm cl} \sim 200$--3000~\Msun, and are highly pressured ($P/k \sim 3\times 10^{6}$~\cmt~K). These states are maintained by a hot ($\sim$$10^{6}$~K), diffuse surrounding medium of equal pressure, supported by the influence of a large number of SN shocks. In order to maintain the observed level of ionization, \citeauthor{lord96} found the average separation of the clouds and ionizing sources had to be of order their size (1--7~pc). This agrees well with the average cluster separations catalogued by \citet{melo05}, and with high-resolution CO observations \citep{mao00, weis01} that also show that the clouds are partly disrupted and dissociated. By modelling the observed forbidden-line profiles, \citeauthor{lord96}~argued that the clouds are clustered in two `hot-spots' of size $125\times 125$~pc located at $\sim$$\pm 15''$ either side of the nucleus, representing a cross-section through the molecular and ionized gas torus found to surround the nuclear bar \citep{larkin94, shen95, achtermann95, weis01}.

%superwind
M82 exhibits one of the largest optically visible outflows or `superwinds' in the local Universe. The outflow is centred on regions A and C \citep{shopbell98, ohyama02}, and is composed of a complex morphology of loops and filaments. Recent ground- and space-based narrow-band optical imaging have shown that, contrary to previous thought, the M82 outflow is comprised of many channels which seem to point back to the individual starburst clumps \citep[][Gallagher et al., in prep.]{wills99}, suggesting that the wind energy is injected by multiple discrete sources (i.e.\ SSCs), rather than a monolithic starburst region \citep*[see e.g.][]{t-t03}. The study of the ISM environment in which SSCs exist (the starburst core) and from which the superwind is driven, can hope to give insights into the initial conditions for galactic wind formation.

\begin{figure*}
\centering
\plotone{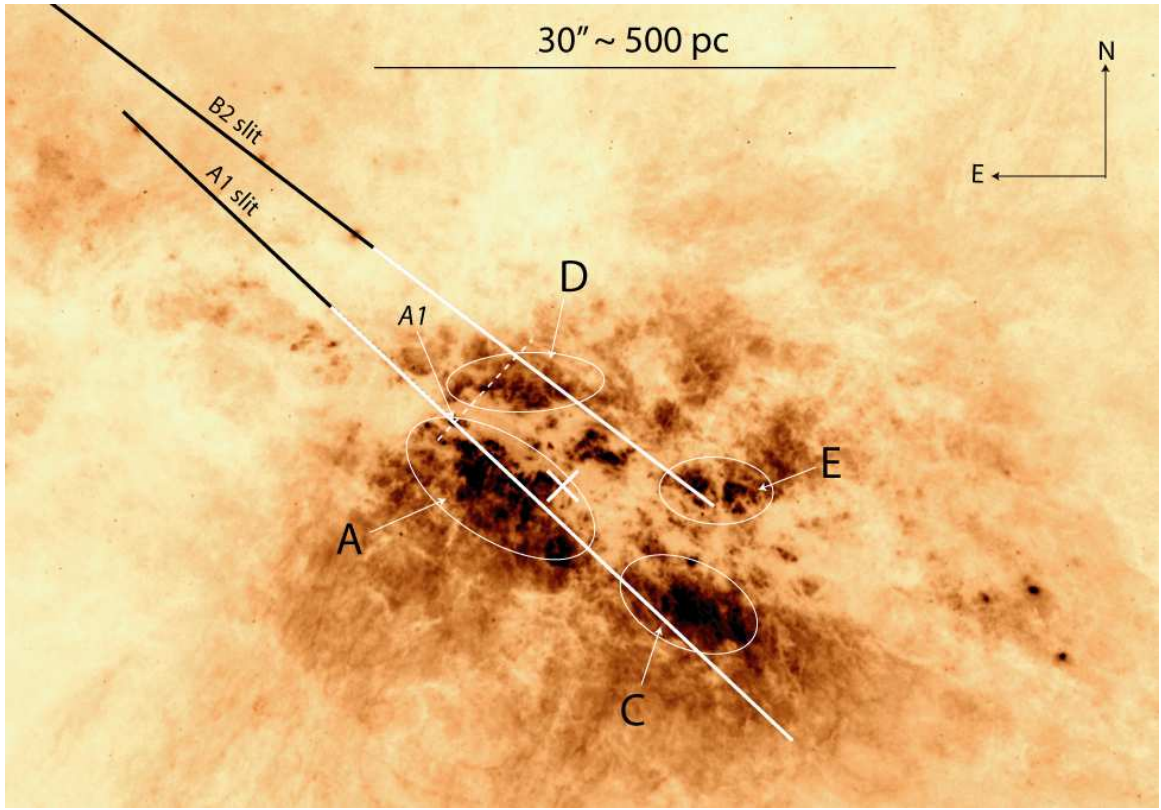}
\caption{\textit{HST}/ACS WFC F658N image with the prominent clumps A, C, D and E identified, together with the location of the SSC M82-A1 and the 2.2 $\mu$m ``nucleus'' (white cross). The location of the two $52\times 0.1$~arcsec STIS slits are also marked, where portions of the 2D slit images corresponding to the white shaded parts of the slits are shown in Fig.~\ref{fig:2d_slit}. The dashed line perpendicular to the slits marks the zero reference point used in the following plots.}
\label{fig:finder}
\end{figure*}

%lead-in remarks
Observations of the ionized gas in the central region of M82 at visible wavelengths with {\it HST} offer the potential of probing the starburst region on sub-arcsecond scales ($0\farcs1 \approx 1.8$~pc) and further quantifying all of the above issues. In this paper, we present spectroscopy obtained with the Space Telescope Imaging Spectrograph (STIS) of the ionized gas near the nucleus, including regions A and C.  We measure the kinematics, densities and extinction of the ionized gas on parsec-scales in order to characterise the nature of the ionized gas in the M82 starburst. 

The remainder of this paper is organised as follows: in Section~\ref{sect:data}, we describe the data and the reduction methods employed, including a description of how we extracted spectra at regular intervals along both slits and fitted the resulting emission-line profiles. The results of our line fitting are described in Section~\ref{sect:results}, together with a discussion of the kinematics and line dynamics. In Section~\ref{sect:prop_gas} we present our analysis of the properties of the ionized gas, including our flux, extinction and electron density results; in Section~\ref{sect:state_ISM}, we go on to describe the physical state of the gas and compare our results to previous models of the ISM. In Section~\ref{sect:SB_struct} we analyse the structure of the starburst and derive a new model for the orientation of the bar and disc with respect to the starburst clumps. In this section we also discuss what implications our results have for our understanding of the superwind. Our findings are summarised in Section~\ref{sect:summary}.

%%%%%%%%%%%%%%%%%%%%%%%%%%%%%%%
\section{Observations and Analysis} \label{sect:data}

\begin{figure*}
\centering
\plotone{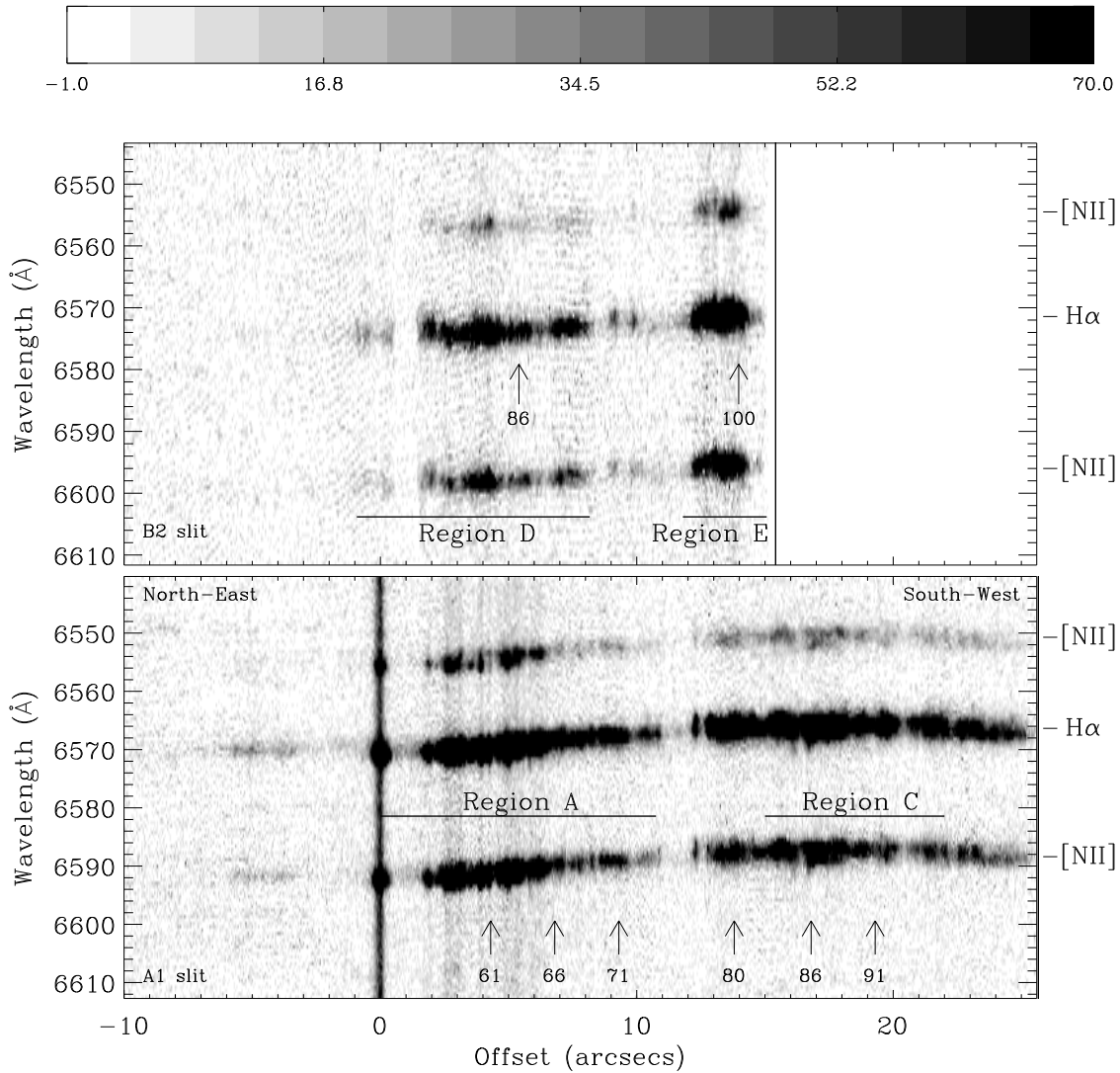}
\caption{STIS G750M two-dimensional spectral images showing the H$\alpha$ and [N\two] emission lines over the portion of the slits covering the nuclear regions (equivalent to the white shaded parts of the slits shown in Fig.~\ref{fig:finder}). The lower plot shows the image from the A1 slit, and the upper plot shows the B2 slit. In both cases, the $x$-scale is in arcseconds offset from cluster A1 (for reference, this point is marked with a dashed line on Fig.~\ref{fig:finder}) and the brightness scale is in units of 10$^{-16}$ erg cm$^{-2}$ s$^{-1}$ \AA$^{-1}$ arcsec$^{-2}$. The emission lines are identified, and the slit orientation is marked. Spectra were extracted using 10 and 20 pixel apertures along the length of both slits (see text); arrows marking particular spectra show the position of the H$\alpha$ profiles shown in Fig.~\ref{fig:ap10_hafits} (where the numbers correspond to the plot titles).}
\label{fig:2d_slit}
\end{figure*}

\begin{figure*}
\centering
\plotone{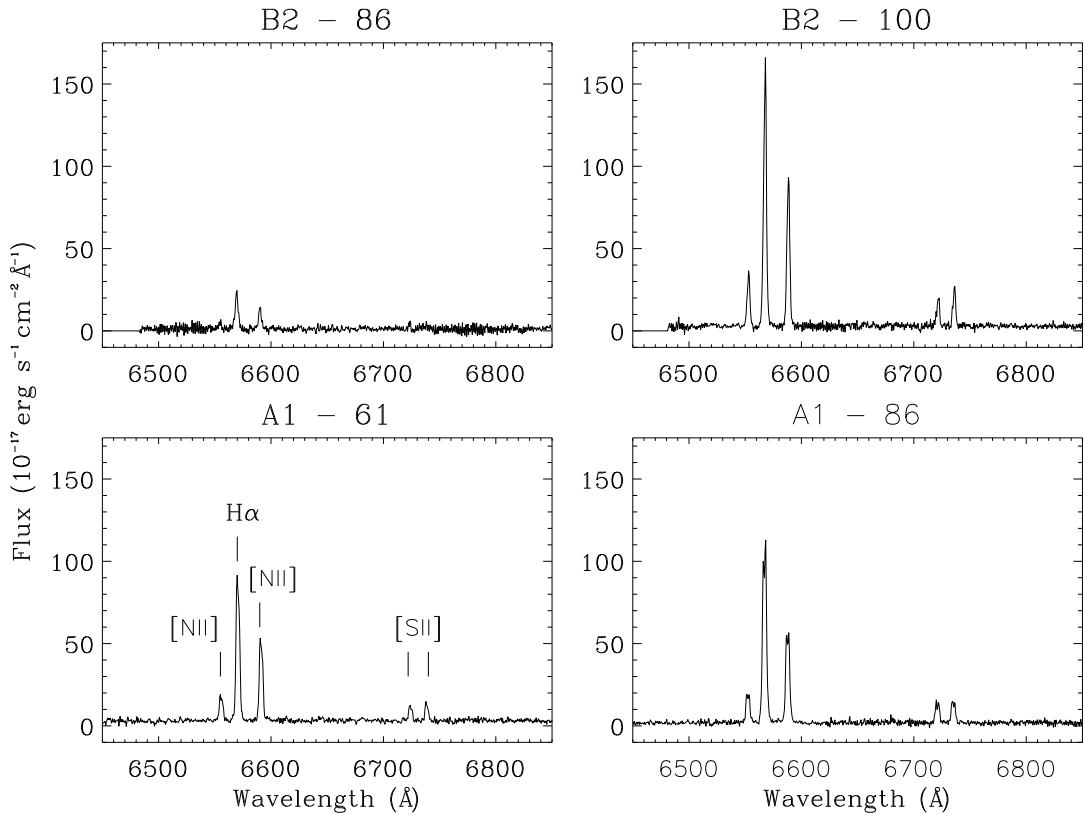}
\caption{Representative STIS G750M spectra of regions A, C, D and E. The numbers in the plot titles correspond to the arrows in Fig.~\ref{fig:2d_slit}. The emission lines are identified on the lower-left plot only.}
\label{fig:spec_plots}
\end{figure*}

\begin{figure*}
\centering
\begin{minipage}{7.5cm}
\includegraphics[width=7.5cm]{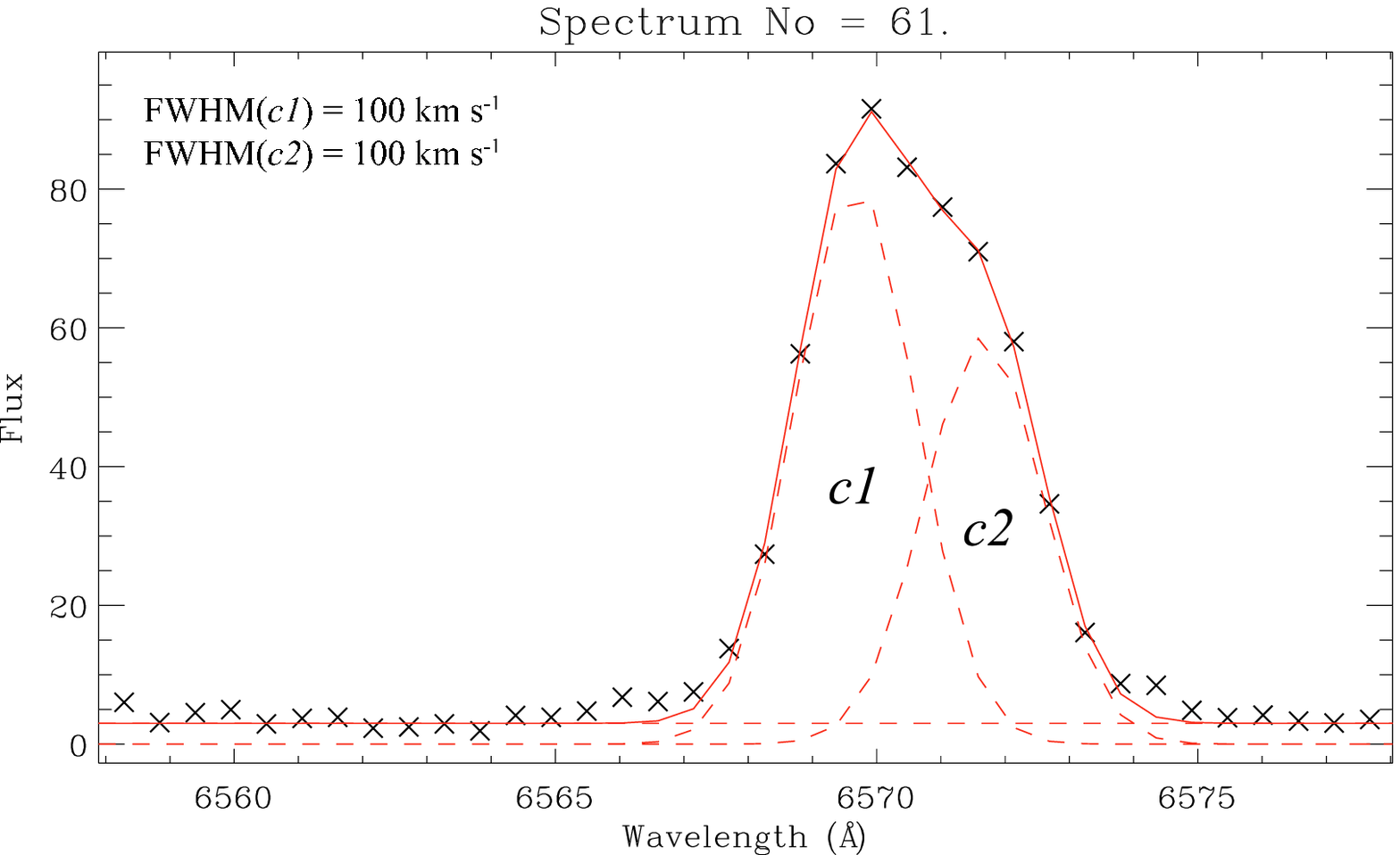}
\end{minipage}
\hspace*{0.8cm}
\begin{minipage}{7.5cm}
\includegraphics[width=7.5cm]{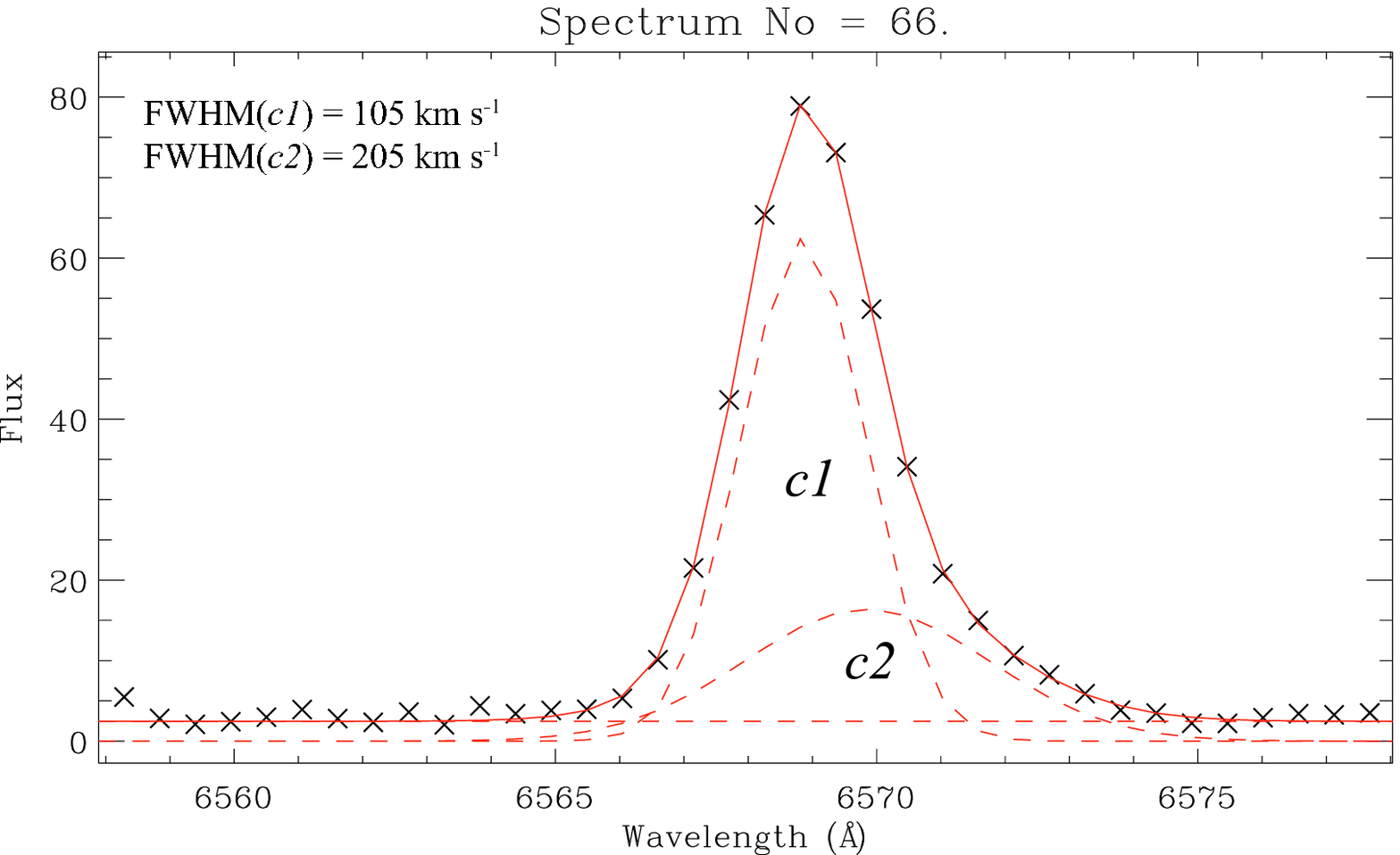}
\end{minipage}
\begin{minipage}{7.5cm}
\vspace{1.2cm}
\includegraphics[width=7.5cm]{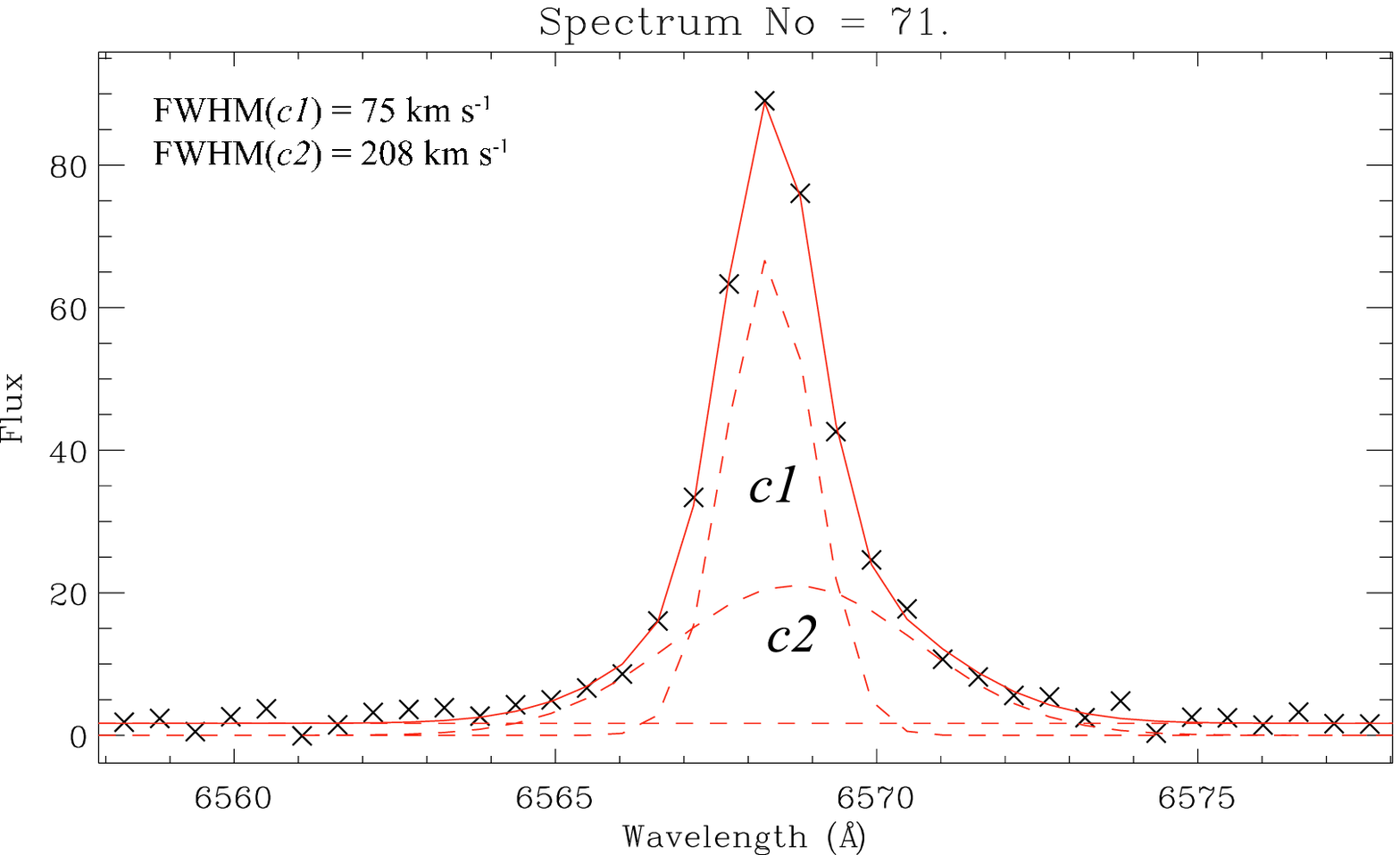}
\end{minipage}
\hspace*{0.8cm}
\begin{minipage}{7.5cm}
\vspace{1.2cm}
\includegraphics[width=7.5cm]{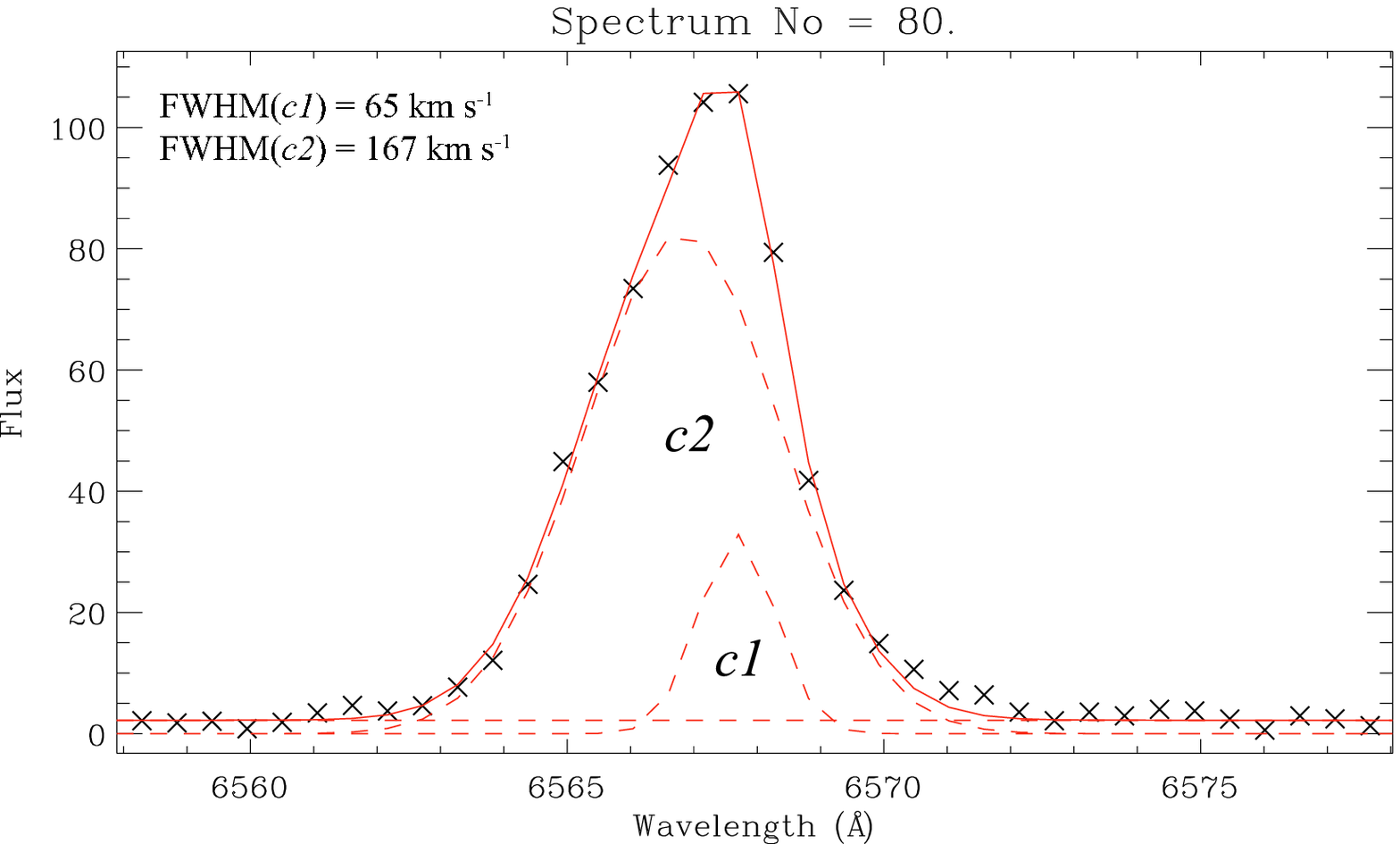}
\end{minipage}
\begin{minipage}{7.5cm}
\vspace{1.2cm}
\includegraphics[width=7.5cm]{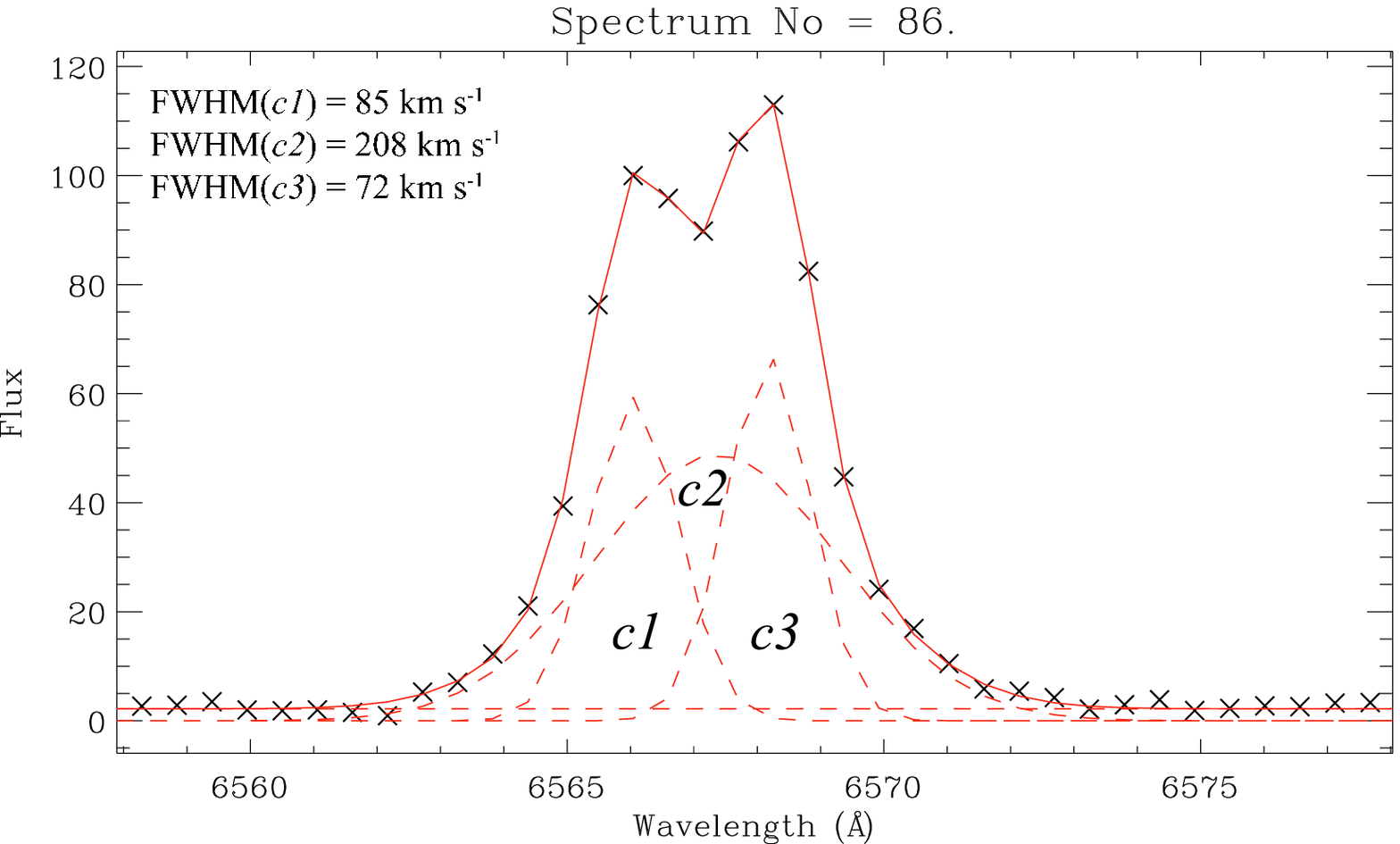}
\end{minipage}
\hspace*{0.8cm}
\begin{minipage}{7.5cm}
\vspace{1.2cm}
\includegraphics[width=7.5cm]{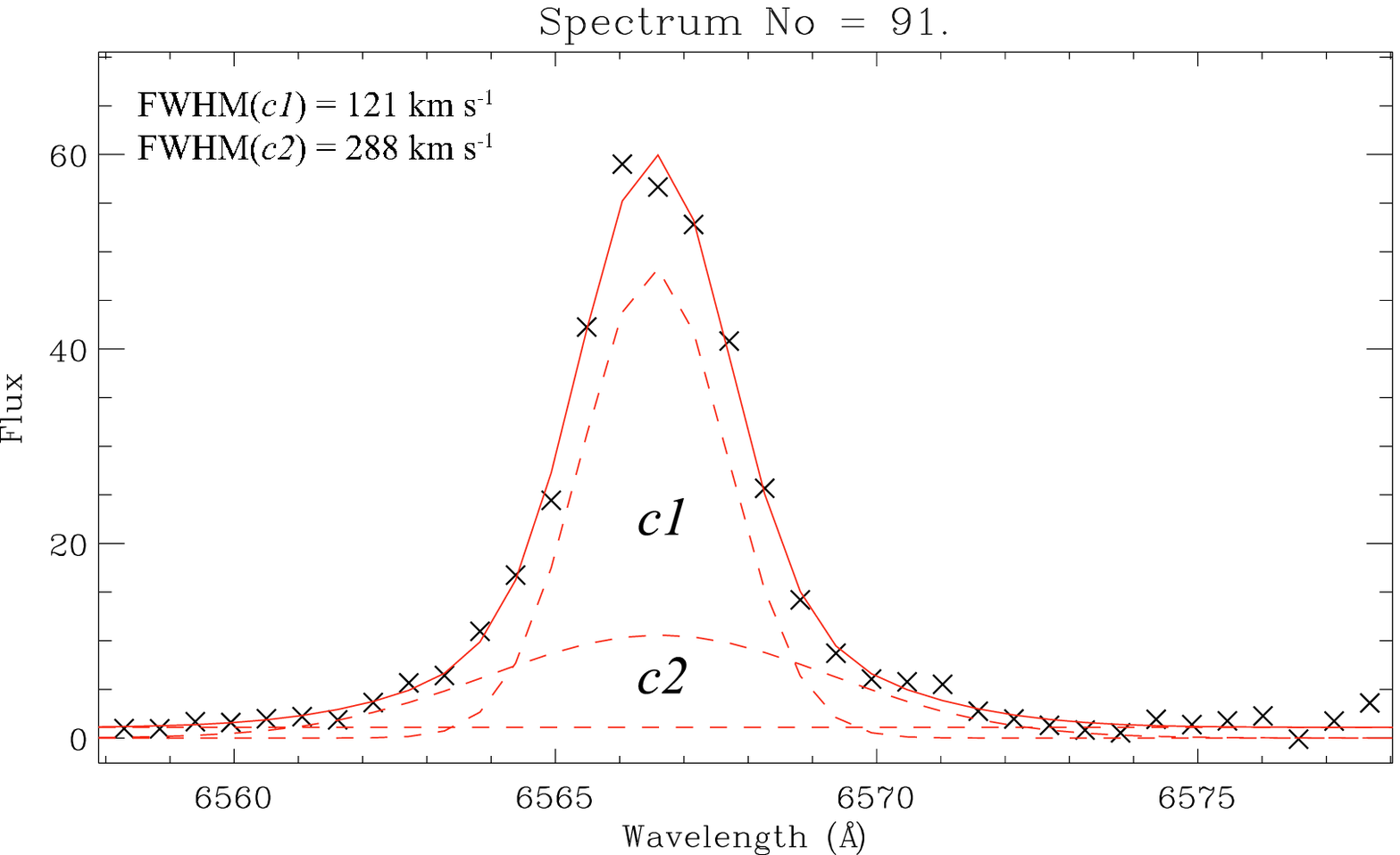}
\end{minipage}
\begin{minipage}{7.5cm}
\vspace*{2.2cm}
\includegraphics[width=7.5cm]{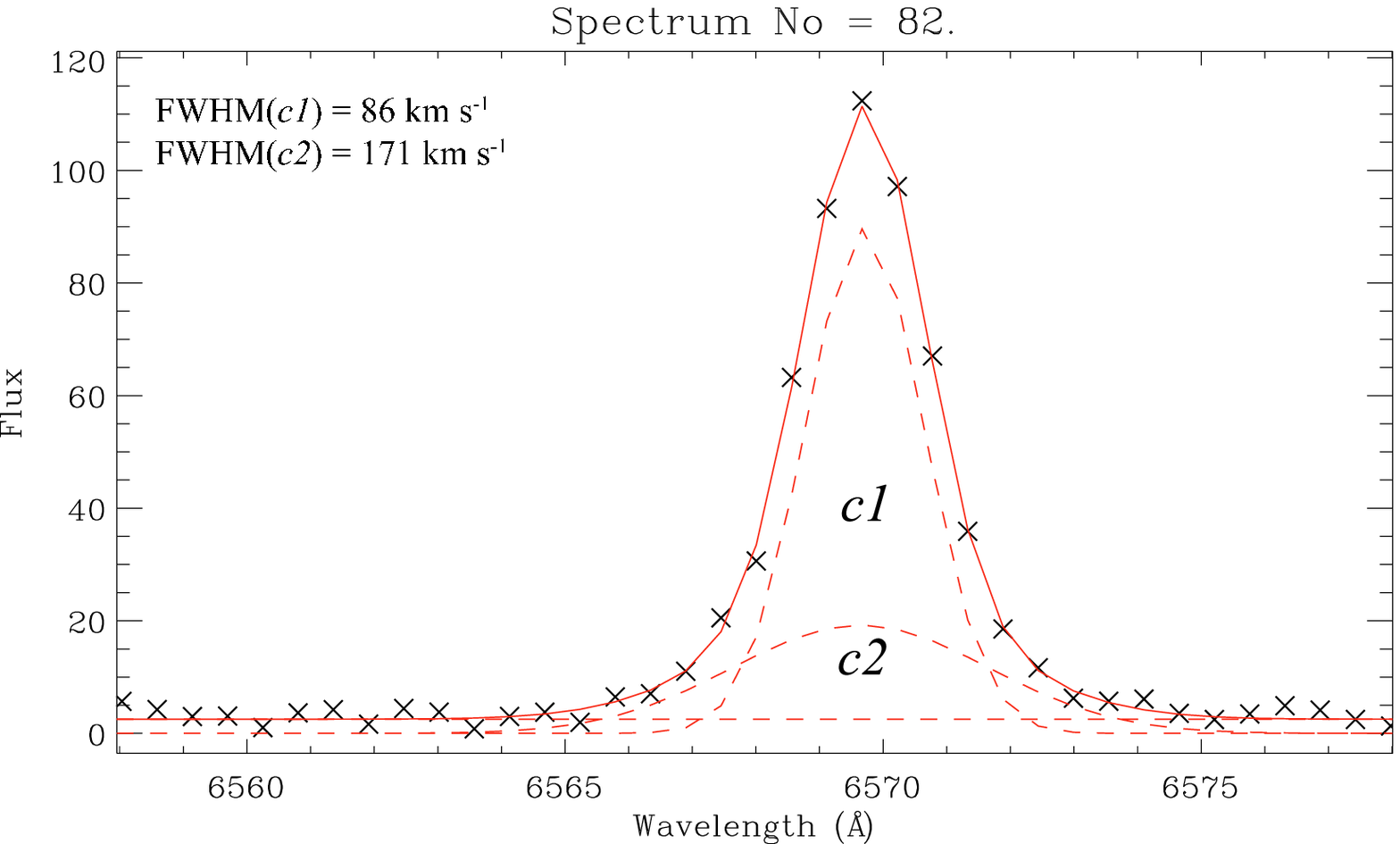}
\end{minipage}
\hspace*{0.8cm}
\begin{minipage}{7.5cm}
\vspace*{2.2cm}
\includegraphics[width=7.5cm]{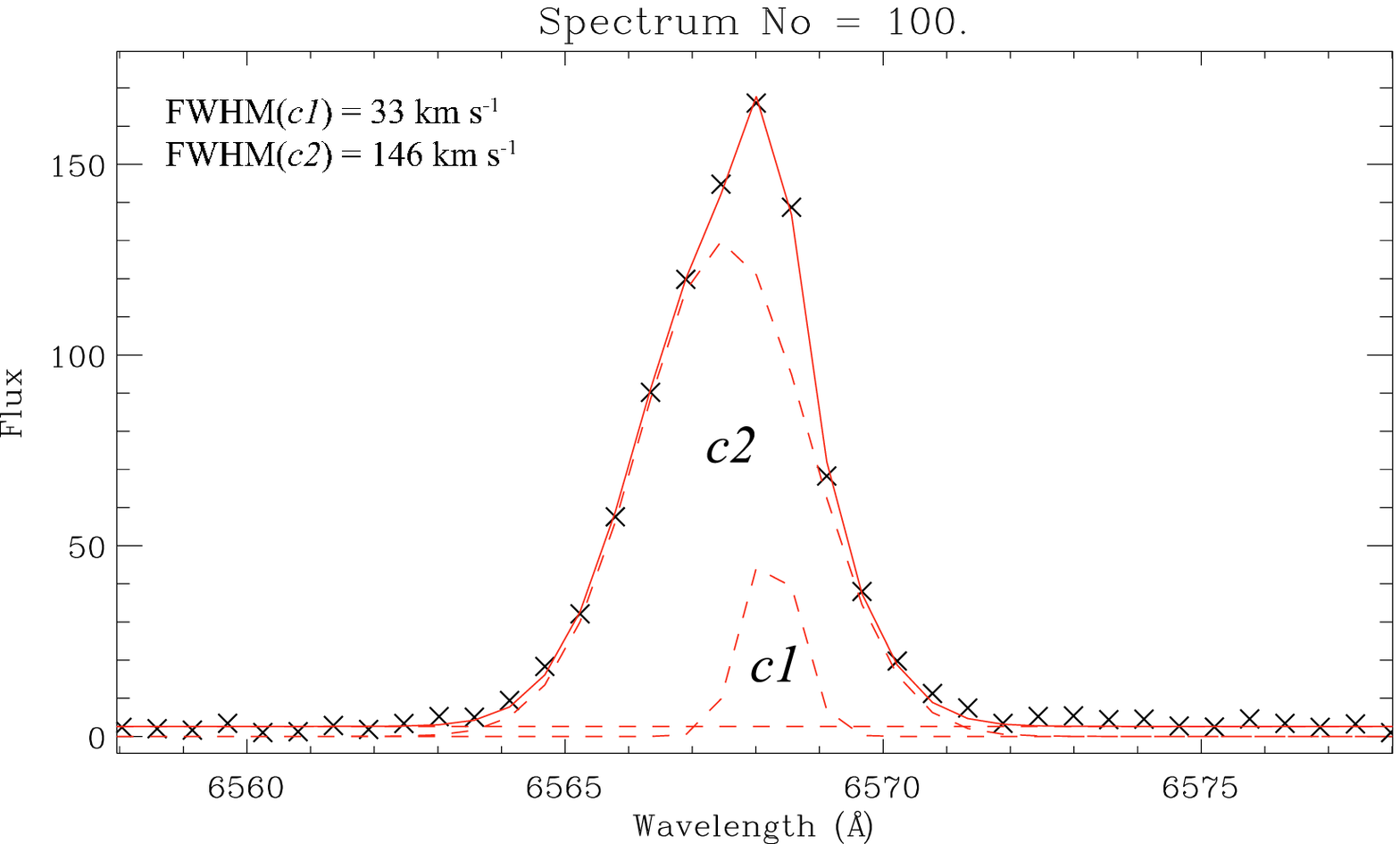}
\end{minipage}
\vspace{0.8cm}
\caption{Example H$\alpha$ line profiles for slit A1 (upper six) and slit B2 (lower two): arrows on Fig.~\ref{fig:2d_slit} indicate where the spectra are located along the slits (numbers correspond to plot titles). This sample was chosen to represent a range in line profile shapes including broad, underlying features and second narrow peaks. The observed data are plotted with un-joined crosses; the flat continuum and individual Gaussian profiles are plotted as dashed red lines with $y$-offsets relative to the zero level, and the summed model profile is plotted with a solid red line. The individual line components of each fit are labelled and their FWHMs are shown. Fluxes are in units of $10^{-17}$~erg~s$^{-1}$~cm$^{-2}$~\AA$^{-1}$.}
\label{fig:ap10_hafits}
\end{figure*}

As part of a larger programme devoted to spectroscopy of SSCs in M82 (GO 9117; P.I.\ O'Connell), we obtained \textit{HST}/STIS spectra for two slit positions crossing the galaxy's starburst core. Fig.~\ref{fig:finder} shows the position of the two STIS slits overlaid on an \textit{HST}/ACS WFC F658N image (GO 9788; P.I.\ Ho). The slits ($52\times 0.1$~arcsecs) pass close to the ``nucleus'' (2.2~$\mu$m peak) and just above and below the dust lane that perpendicularly bisects the disc of M82. The first slit, which we will refer to as slit A1, intersects the core of region A and the southern edge of region C, and was centred on the brightest isolated cluster in region A (which we have designated M82-A1) with position angle PA = $229^\circ$. The second slit, slit B2, was positioned on a bright cluster \citepalias[B2-1;][]{smith06} towards the north-east of the starburst core with a position angle PA = $235^\circ$, and passes through portions of regions D and E. The coordinates and exposure times for each slit position are given in \citetalias{smith06}.

\subsection{Description of 2D spectra} \label{sect:desc_spectra}
Spectra were taken with two gratings giving coverage from 2900--5700\,\AA\ with the G430L and 6295--6865\,\AA\ with the G750M setting with a spatial resolution of $0.05$~arcsec~pixel$^{-1}$. A detailed description of the data reduction procedure is given in \citetalias{smith06}. The spectral resolution of the data was measured for the G430L data by taking the average FWHM of Gaussian fits to the unresolved H$\beta$ line at various points along the length of the slit where extended H$\beta$ emission is observed. We found the resolution to be 2.5~pixels (in good agreement with the quoted value of 2--3~pixels given in the STIS Data Handbook for extended objects), corresponding to 6.7~\AA\ in the G430L grating and 1.4~\AA\ in the G750M grating. Since variations in slit illumination and of the source structure at each point could affect the resolution slightly, it is possible that some of the narrowest lines we measure (uncorrected FWHM~$\le$~75~\kms) may be unresolved. The material covered by the majority of the north-easternmost portion of both slits is very faint (see Fig.~\ref{fig:finder}), and not detected in our observations. The extinction in this region as measured by \citet{satyapal95} and \citet{alonso03} may be very high, A$_{V} \sim 10$--15~mag. In spectra where we do detect emission, we observe the nebular emission lines of H$\beta$, [O\three]$\lambda 5007$, [N\two]$\lambda\lambda 6548,6583$, H$\alpha$ and [S\two]$\lambda\lambda 6716,6731$.

Fig.~\ref{fig:2d_slit} shows the H$\alpha$ and [N\two] spectral region of the G750M 2D spectral images for slits A1 and B2. Both 2D slit images are able to share the same $x$-axis since their position angles (PAs) are sufficiently similar (scale is given in arcseconds offset from the position of M82-A1 -- marked with a dashed line on Fig.~\ref{fig:finder}). The spatial extent of the slit images is also restricted to show only the nuclear regions where emission is actually detected (represented by the white shaded section of the slits in Fig.~\ref{fig:finder}). We see emission arising from both discrete sources and holes in the foreground dust screen. Bright emission from the ambient interstellar gas is also present.

Fig.~\ref{fig:2d_slit} clearly shows how the emission line profiles vary continuously along the length of the slit. It would be ideal to track these variations on a pixel-to-pixel basis, but in order to make accurate measurements of the gas conditions, the spectra need to have a sufficiently high signal-to-noise (S/N) ratio. Therefore we summed the spectra using multiple-pixel apertures. To measure the kinematics of the gas, observations of a strong emission line (e.g.~H$\alpha$) are all that is required. However, to derive gas properties that require information from additional, fainter lines, a higher S/N was needed. To satisfy these two criteria, we extracted two sets of spectra from each of the 2D images. A high S/N set for each slit was made by stepping along in 20 pixel intervals ($\equiv$\,$1\farcs0$), extracting apertures of width equal to this interval (hereafter referred to as set A20), and resulted in spectra from slit A1 with an average S/N of 10 in H$\beta$ (G430L grating) and 20 in H$\alpha$ (G750M grating). In order to maximise spatial-resolution for the kinematics measurements whilst maintaining sufficient S/N for accurate line fitting, we decreased the extraction widths to 10 pixels ($\equiv$\,$0\farcs5$) separated by intervals of 10 pixels to create the second data-set (hereafter referred to as set A10). Extracted spectra were given sequential IDs starting from the north-east of each slit to identify them in later analysis. No background subtraction was attempted for either of the slit A1 or B2 datasets due to the bright and variable nature of the diffuse emission in the starburst core.

Fig.~\ref{fig:spec_plots} shows four representative spectra from regions A, C, D and E illustrating the variation between these clumps. The arrows on Fig.~\ref{fig:2d_slit} mark the position that each spectrum shown in Fig.~\ref{fig:spec_plots} was extracted from, where the numbers indicate the spectrum IDs referred to above, and correspond to the individual plot titles in Fig.~\ref{fig:spec_plots}.

\subsection{Cluster distributions and approximate ages within the clumps} \label{sect:ews}
The clumps of emission seen in Figs~\ref{fig:finder} and \ref{fig:2d_slit} represent a densely packed mixture of individual clusters and their surrounding H\two\ regions, ionized gas clumps and bright diffuse ISM seen through holes in the foreground dust screen, all with sizes and separations on the order of a few to a few tens of pc. The extent to which these phenomena dominate the physical nature of the spectra along the length of the slits varies dramatically.

In the cores of the starburst clumps, the slits cross a number of individual clusters (see e.g.~\citetalias{smith06}, figure 1 inset). We have experimented with extracting spectra close to individual sources catalogued by \citet{melo05}, but unfortunately the slits to do not pass directly through the centre of any isolated clusters, making the extraction of unique spectra difficult. Those that we did extract were of too low S/N to allow accurate model fitting. Measuring the equivalent width (EW) of the H$\alpha$ line in these core regions can, however, give an approximate indication of age trends in the stellar population by comparing to evolutionary synthesis models. It must be borne in mind, however, that since there is no way that we can measure the light from individual clusters, effects such as crowding, slit losses, and the presence of bright ambient diffuse emission will strongly affect the meaning of our EW measurements.

By measuring the H$\alpha$ EW along the central core regions of both slits and using the EW predictions of a {\sc starburst99} \citep[{\sc sb99};][]{leitherer99}, $10^6$~\Msun\ model with a Kroupa IMF formulation, we find that the cluster ages in regions A, C, D and E lie between 5.5 and 6.5~Myr (log(EW) $\approx$ 1.5--2.5). If, however, a significant fraction of the Lyman continuum photons are absorbed by dust \citep[][and discussed in Section~\ref{sect:emission_meas}]{dopita02, dopita06a}, then cluster ages determined in this way will be overestimated. Nevertheless, these values are consistent with previous age estimates of clusters in the starburst core \citep{forster03, smith06}.

\subsection{Line Profile Fitting} \label{sect:fitting}
The S/N and spectral resolution of the 10-pixel wide spectra are high enough to resolve multiple components in H$\alpha$ in many regions along the slit. In general, a bright, narrow component is superposed on an underlying broad, fainter component, but obviously the identification of the broad component is dependent on the S/N of the particular spectrum. The component profiles are best approximated by a Gaussian function since the line broadening mechanism is dominated by Doppler effects caused by the gas temperature or turbulent state.

Examples of H$\alpha$ line profiles from different parts of regions A, C, D and E are shown in Fig.~\ref{fig:ap10_hafits} together with the Gaussian profiles needed to model the integrated shapes (see below). Again, the position along the slit that each H$\alpha$ profile has been extracted from is marked by numbered arrows on Fig.~\ref{fig:2d_slit}. These were chosen to represent a range in the variety of line shapes seen: bright, narrow emission lines with underlying broad components; double-peaked lines with no detectable underlying broad emission; and mixtures of the two. Where a broad component is present, it is often, but not always, redshifted compared to the narrower component. Double-peaked lines with velocity differences of $\sim$50--100~\kms\ are seen in both the cores of region A and C.

We have used the general-purpose IDL-based curve-fitting utility, \textsc{pan} \citep[Peak ANalysis;][]{dimeo}, to automate the Gaussian fitting process of each of the extracted spectral lines. Briefly, \textsc{pan} embodies an interactive environment for specification of a profile's initial guess parameters, and uses a $\chi^{2}$ minimisation algorithm to optimise the fit to the data. A more detailed description of the program and how we have customised it for our use is given in \nocite{westm07a}Westmoquette et al.\ (2007a).

Each line profile was fit with the appropriate number of Gaussian components, giving a measurement of the radial velocity, flux, and FWHM for each component. The lowest S/N lines could only be fit with a single Gaussian. However along most of the slit where emission is detected, the line profile shape distinctly contains more than one component. In cases where a double-component fit was most appropriate (determined from a comparison of the $\chi^{2}$ fit value to the single component result, and by visual inspection), the initial guess was always made setting the first Gaussian as the narrow component (hereafter referred to as \textit{c1}), and the second Gaussian as the broader component (hereafter \textit{c2}); in the case of a double peak, the bluest component was assigned as \textit{c1} and the redder one as \textit{c2}. For a few cases a triple-component fit was necessary, and the additional component (\textit{c3}) was always specified with a initial guess assigning it to a supplemental narrow line at the same wavelength as the main narrow line. This consistent approach helped limit the confusion that might arise during analysis as to which Gaussian fit belonged to which component of the line, as well as aiding the $\chi^{2}$ minimisation process employed by \textsc{pan}. This assignment convention means that wherever an underlying broad component exists, we always refer to it as \textit{c2}.

%%%%%%%%%%%%%%%%%%%%%%%%%%%%%%%
\section{Kinematics of the ionized gas} \label{sect:results}

There have been a number of previous studies aimed at measuring the characteristics of the gaseous component in the disc and wind of M82 \citep[e.g.][]{gotz90, heckman90, mckeith93, mckeith95, shopbell98}. From these studies, it is apparent that the wind outflow, as defined by two well-separated ($\sim$300~\kms) velocity components, is observed out to distances greater than 1~kpc from the nucleus along the minor axis. The disc of M82 is inclined at an angle of $\sim$80$^{\circ}$ \citep{lynds63} such that the near side of the disc is projected on to the north-west side of the nucleus and the nuclear regions, including clumps A, C, D and E, are observed from underneath through the southern side \citep[although this has been disputed in the past;][]{wills00}. We therefore directly view the roots of the wind near regions A and C. The two STIS slits are inside the 300~pc wind injection zone \citep{shopbell98} where disc material is being entrained into the flow, and likely sample much of the inter-cluster material within the main starburst clumps.

\subsection{Radial velocity variations} \label{sect:results_radvel}

\begin{figure*}
\centering
\plotone{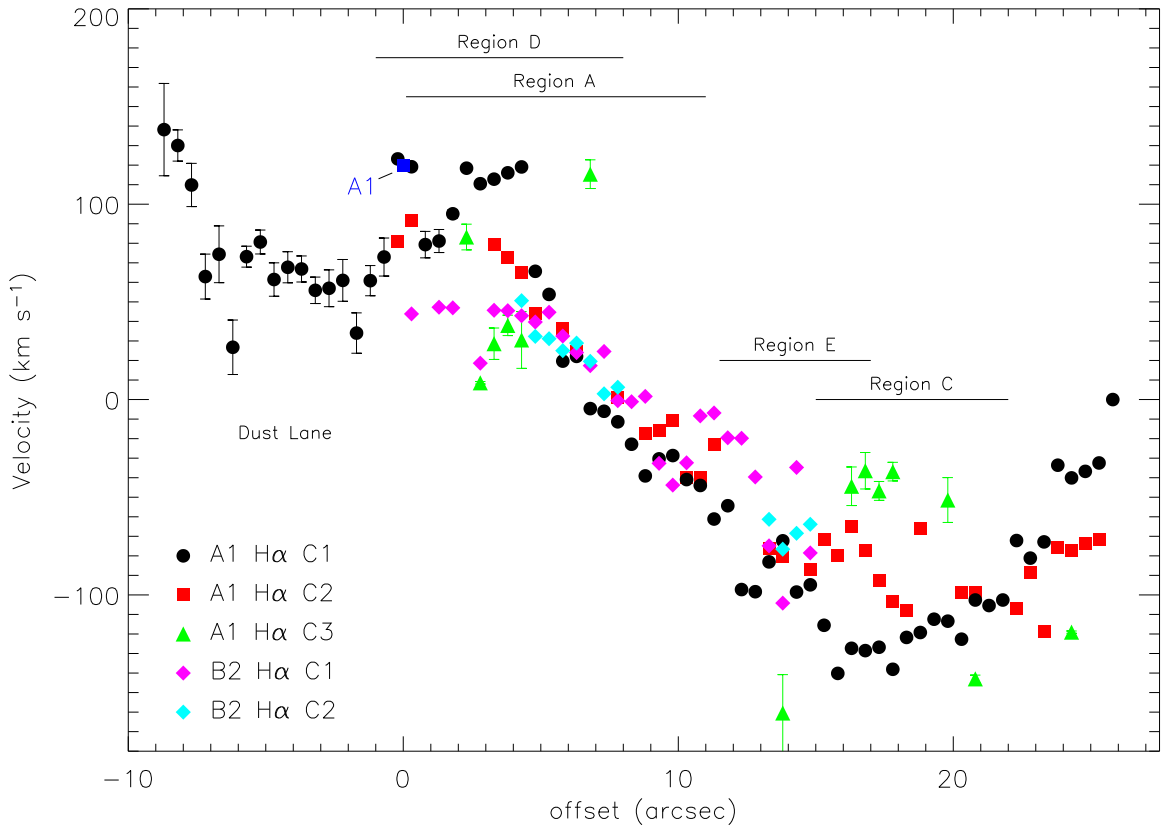}
\caption{Heliocentric H$\alpha$ radial velocities measured from the A10 spectrum sets, in units of \kms\ with respect to the systemic velocity of M82 ($v_{\rm sys} = +200$~\kms); error bars are not shown because they are approximately the same size as the symbols used. The $x$-axis scale corresponding to both slits is measured in arcseconds offset from the position of M82-A1. The projected distance along the slit from the location of cluster A1 to the nucleus is $\sim$5$''$. Measurements from both slits are shown on the same plot for ease of comparison; the approximate extent of regions A and C that coincide with slit A1, and regions D and E that coincide with slit B2 are marked. \textit{c1} refers to the bright, narrow component, \textit{c2} to the broad component, and \textit{c3} to the longer wavelength component when the line is resolved into two narrow components.}
\label{fig:ap10vel}
\end{figure*}

The measured radial velocities of the H$\alpha$ emission-line components from Gaussian fits to the A10 spectrum-sets are shown in Fig.~\ref{fig:ap10vel}, and are plotted with respect to the systemic velocity of M82 \citep[$v_{\rm sys} = +200$~\kms;][]{vaucouleurs91, mckeith93}. We have again plotted the datasets from both slits together. As described in the above section, in general the H$\alpha$ line profile in the brightest parts of all four regions is composed of a bright, narrow component superimposed on a broad, fainter component with a FWHM $\sim$ 100~\kms. At a few positions, we see a third component (\textit{c3}). Error bars are shown for points to the north-east of M82-A1 and for all \textit{c3} detections. For most of the other points, the error bars are smaller than the size of the plotting symbol.

The brightest component, \textit{c1}, is expected to trace the velocity of the densest (i.e.~bulk of the) ionized gas in M82. From Fig.~\ref{fig:ap10vel}, the slit A1 velocity curve for \textit{c1} reaches a maximum velocity of $\approx$+100~\kms\ at an offset of $\sim$$+2''$, flattens out between $-2''$ to $-7''$ coincident with the conspicuous dust-lane bisecting the galaxy disc \citep[see Fig.~\ref{fig:finder}; also][]{oconnell95}, then begins to rise again beyond $-7''$. The velocity of M82-A1 \citepalias{smith06}, together with two points either side of it, are redshifted by $\sim$40~\kms{} from the general trend. On the opposite side, the rotation curve shown by \textit{c1} reaches a maximum blueshifted velocity of $\approx$$-130$~\kms\ then begins to decrease again beyond $+20''$. At these offsets, however, the A2 slit is farthest from the major-axis and begins to sample the inner regions of the wind -- we return to this point below. We measure a \textit{c1} H$\alpha$ velocity gradient for the region nearest to the nucleus of 12~\kms{}~arcsec$^{-1}$, in excellent agreement with the measurements of \citet{mckeith93} and \citet{shopbell98} of 11~\kms~arcsec$^{-1}$. The radial velocities of the \textit{c1} gas in slit B2 follow the same velocity trend as the gas sampled along slit A1, but with a shallower gradient of 7~\kms~arcsec$^{-1}$. We discuss what these measurements tell us about the rotation curve in Section~\ref{sect:SB_struct}.

The broad underlying component, \textit{c2}, largely follows the \textit{c1} velocity curve although there are some deviations. In the core of region C, \textit{c2} is redshifted by up to 60~\kms\ compared to \textit{c1}. Here we detect a third component (\textit{c3}), which is redshifted by a further 30--40~\kms, and may represent the front side of an expanding structure and possibly the roots of the wind flow. Further along the slit, to the south-west of region C, \textit{c2} becomes blueshifted by $\sim$40~\kms\ with respect to \textit{c1}, and there are two detections of a 40--80~\kms{} blueshifted third component. Again, these could be signatures of expanding structures and/or ordered flows. An unusual situation occurs at $\sim$$+4''$ offset (right in the core of region A), where \textit{c1} is redshifted compared to its expected velocity, and it is \textit{c2} that instead conforms to the trend. Here we also detect a third component, blueshifted by $\sim$80~\kms\ compared to \textit{c1}, which might explain the unexpected velocities if they are both two halves of an expanding shell where the faintest component is the nearest. In slit B2, we can only detect \textit{c2} in the brightest parts of regions D and E, and find that their velocities are similar in all cases to that of \textit{c1}, except in the core of region E where there is evidence of a 20--40~\kms{} offset. Critically, however, none of the radial velocity differences measured equal the galactic wind velocity shifts of $\sim$300~\kms. This evidence suggests that we are observing inside the injection zone where organized flow velocities are low \citep[consistent with][]{shopbell98} and/or that the flows at this point are oriented in the plane of the sky (i.e.~in a poleward direction). 

\subsection{Line widths and components} \label{sect:results_fwhm}

\begin{figure*}
\centering
\plotone{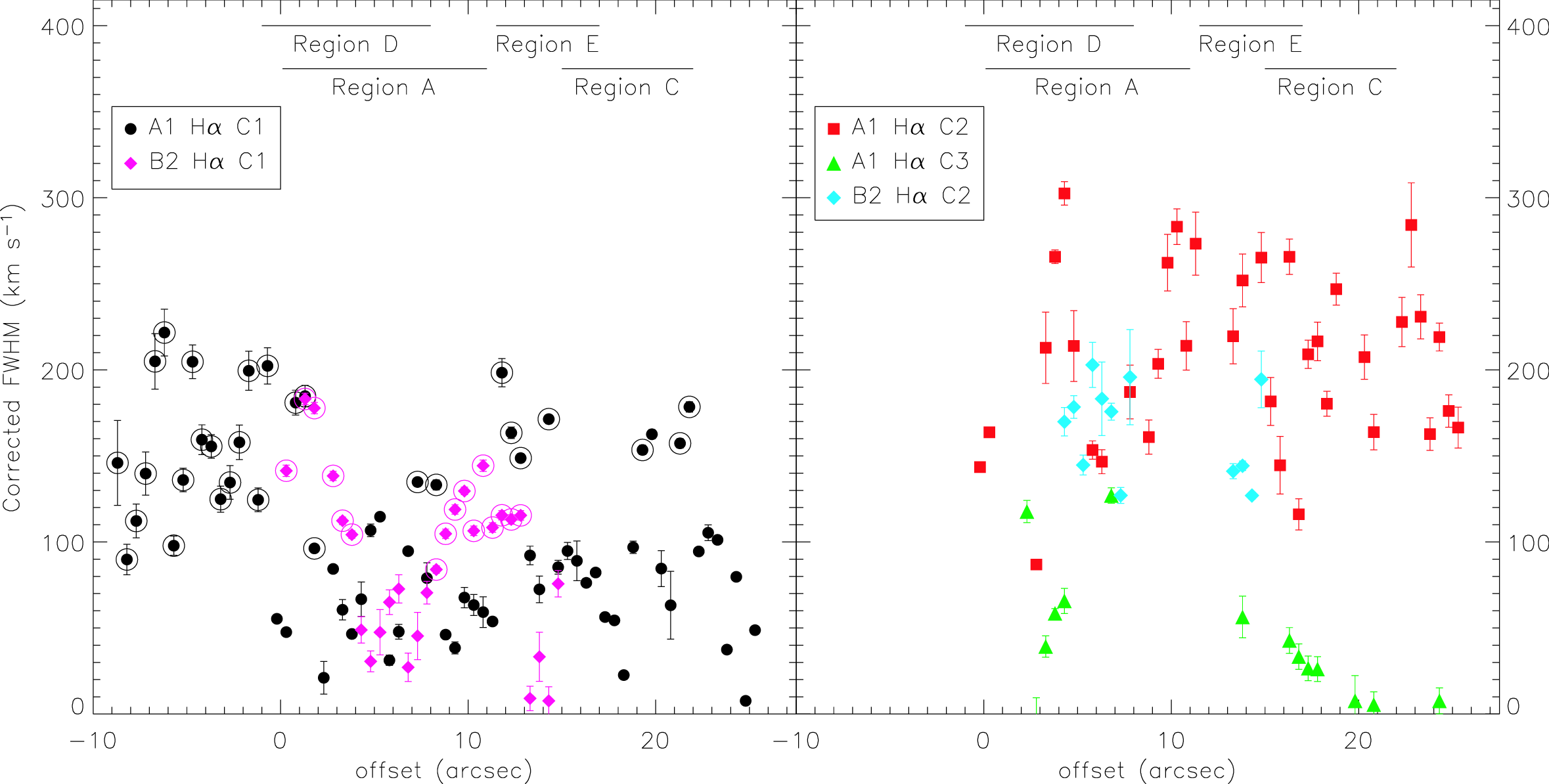}
\caption{H$\alpha$ FWHMs measured from the A10 spectrum sets for \textit{c1} (left-hand panel) and \textit{c2} \& \textit{c3} (right-hand panel). The units are in \kms\ corrected for instrumental broadening, and the $x$-axis scales are measured in arcseconds offset from the position of M82-A1. Circled points are those where no more than one component was required to fit to the line.}
\label{fig:ap10fw}
\end{figure*}

Fig.~\ref{fig:ap10fw} shows the FWHM of the individual H$\alpha$ line components for both slits, corrected for the instrumental resolution (1.4~\AA) but not thermal broadening. As mentioned in Section~\ref{sect:data}, small variations in the spectral resolution due to changes in the slit illumination/source structure could mean that the narrowest lines are only marginally resolved. The large scatter in \textit{c1} FWHM values (left-hand panel) across the two slits can be partly explained by a contribution from multiple unresolved kinematic components. We have attempted to show this by highlighting \textit{c1} profiles from both slits that only required a single Gaussian to fit with circled symbols: the FWHMs of circled points are, in almost every case, broader than the narrow component (\textit{c1}) of a multiple-Gaussian fit. This is particularly obvious in the $-9''$ to $0''$ region of slit A1 where a second component cannot be fit at all due to the low S/N of the extracted spectra. It is very unlikely that these differences in width reflect real changes, so we conclude that multiple components are present everywhere, but the S/N is not sufficient in all cases to permit a more accurate fit to be made. This argument is strengthened by the observation that in every spectrum extracted from either slit where the H$\alpha$ line has a high enough S/N, we see evidence of a \textit{c2}- or \textit{c3}-type component. 

To measure the following averages, we ignore all single-Gaussian-only fits, as these are assumed to be biassed by unresolved components. For slit A1, we find an average FWHM of $\sim$$60\pm 40$~\kms\ for \textit{c1} (whereas the average width of single-Gaussian-only fits is $\sim$$160\pm 40$~\kms). For \textit{c2}, we find an average FWHM of $\sim$$210\pm 60$~\kms, but in a few cases the width of \textit{c2} approaches 300~\kms. Where detected, the mean FWHM of \textit{c3} is $\sim$50~\kms, which is consistent with the mean line width of \textit{c1}. In general, the FWHM of the gas in slit B2 is narrower: in region D the width of \textit{c1} is as low as 20--30~\kms, with a corresponding \textit{c2} FWHM of 150--200~\kms. In parts of region D and the area between regions D and E, only one component can be fit (shown by circled symbols), with a width of $\sim$100--140~\kms. The line widths in region E are similar to region D, and in both regions \textit{c2} is never $>$200~\kms.

\subsubsection{Line broadening mechanisms}
Fig.~\ref{fig:ap10fw} shows that all but a few line components are fit with Gaussians with widths $>$~30~\kms{}. Interestingly, the widths are systematically narrower in regions D and E (even including those that could only be fit with a single component), possibly indicating a more quiescent gas state. We will now explore the possible explanations for the observed widths of the narrow (\textit{c1}) and broad (\textit{c2}) lines in turn.

Supersonic line widths have been observed in many intense star-formation sites both in nearby galaxies (30 Dor: \citealt{chuken94, melnick99}; NGC 604: \citealt{yang96}; NGC 2363: \citealt{roy92, g-d94}) and in more distant dwarf galaxies \citep{izotov96, homeier99, marlowe95, mendez97, sidoli06, rozas06b}. In these studies, gravitational broadening through virial motions of ionized gas clouds \citep{melnick77, terlevich81}, and multiple unresolved kinematical components along the line-of-sight \citep{chuken94} have been proposed as explanations for the broadening of the brightest component (what we refer to as \textit{c1}). A number of other mechanisms proposed, such as dust or electron scattering \citep{roy92}, can be immediately disregarded on physical grounds.

The fact that we rarely measure FWHMs $<$~30~\kms{} across both slits indicates a global origin for a large fraction of the observed widths. A clear lower-limit to the width of \textit{c1} observed in some studies of similar environments \citep{m-t96, martinez07, westm07a, westm07b} supports this explanation. The only two broadening mechanisms that can work on a global scale are gravitationally induced virial motions and the stirring effects of wide-scale, intense star-formation on the ambient ISM. We therefore propose that the turbulent motions induced by these two mechanisms provide the observed base level of line-broadening present over the whole of the region observed. However, neither of these globally determined values can account for the large local variations in \textit{c1} line-width above 30~\kms\ up to 100~\kms. The bright background and variable extinction lends support to the fact that an inevitable additional contribution must result from including multiple, superimposed, small-scale kinematical components along every sight-line, some of which may be of fairly high velocity, within the small-scale emitting regions in the core of M82 (Section~\ref{sect:emission_meas}). We already alluded to this effect above, where we invoked the unresolved nature of the line profile to explain the increased width of the profiles where only a single-Gaussian was required for a satisfactory fit. In fact observations presented by \citet{rozas06b} show that multiple emission line components are very common in H\two\ regions, and can be interpreted as expanding shells driven by the mechanical energy input of SNe.

A ubiquitous broad underlying component (FWHM $\sim$ 150--250~\kms) is observed in both slits where the H$\alpha$ line is of sufficient S/N, which cannot be broken down into individual subcomponents. Like the phenomenon of the supersonic narrow component, broad underlying profiles have also been observed in many intense star-formation sites \citep[see in particular][]{melnick99}, but to our knowledge have never been studied in detail in the central regions of M82. Observations of the broad component in other systems are less well constrained, particularly for more distant galaxies, so the proposed explanations cited in the literature are more varied. These include: rotational effects, direct observations of stellar/cluster winds \citep{g-d94}; the effects of SN remnants \citep[SNRs;][]{roy92, izotov96}; large-scale superbubble expansion and/or blow-out \citep{roy92, marlowe95, t-t97}; multiple resolved/unresolved expanding shells \citep{homeier99, rozas06b}; champagne flows \citep{yorke84, t-t00, melnick99}; and ablation of gas from molecular clumps resulting from the impact of stellar/cluster winds \citep{melnick99}.

Only one of these proposed explanations, that of galactic rotation, is not due to the effects of massive stars, and may be discounted by considering the following: the maximum radial velocity relative to the systemic velocity of M82, $v_{\rm rad}^{\rm max}$, is of the order 100~\kms. If this were purely caused by rotation, then, when looking through one side of the galaxy, the line width would also be $\sim$100~\kms. However, we see line widths of up to 300~\kms, thus proving that the broad line component cannot result from simple rotation. In reality, the $v_{\rm rad}$ observed is due to large-scale orbital motion, not local gas rotation, and rotational broadening is likely close to negligible.

We can also discount direct observations of the stellar winds themselves from the spatial distribution of the line emission. Similarly we are not directly seeing supernova remnants (SNRs) as we do not find extensive evidence for shock spectra. The effects of large-scale superbubble blow-out are highly position dependent and would be expected to produce strong, clearly observable line-splitting, thus not making this a suitable explanation for the observed line broadening. Although multiple unresolved shells are very likely to contribute to the line broadening of \textit{c1}, it is difficult to understand how multiple discrete kinematical components could produce a smooth, Gaussian shaped, 200--300~\kms\ wide profile with no evidence of the discrete components themselves \citep[see also][]{melnick99}. Champagne flows are not expected to produce outflow speeds much higher than the sound speed in the photoionized gas (10~\kms\ for gas of $T\sim 10^{4}$~K), and are therefore not capable of producing the level of broadening we observe. Furthermore, many of the previous studies for which these mechanisms were proposed are of much larger physical regions than what we are observing, meaning that they may not be that appropriate to our $\sim$10~pc scale observations. This leaves only the final explanation: that of hydrodynamical evaporation/ablation of gas from the surface of illuminated gas clouds.

These arguments are strongly supported by results from an analysis of spatially-resolved spectroscopy of a similar, high-energy environment in the starburst galaxy NGC 1569, presented in \nocite{westm07a, westm07b}Westmoquette et al. (2007a, 2007b). Here, we find striking spatial correlations between the width of \textit{c2} and the intensity of \textit{c1}, which strongly supports the hydrodynamic interaction mechanism.

To explain this mechanism, we can envision a situation whereby the energetics and dynamics of the brightest clumps are driven by an interaction with the high-energy, ionizing photons and fast-flowing wind from the surrounding star clusters, and that all the observed emission originates from a thin interface layer at the clump surfaces. As the gas clouds are impacted by hot, fast, cluster winds, thermal evaporation and/or mechanical ablation of gas from their surfaces produces a highly turbulent velocity field that pervades the whole region as the gas becomes entrained into the flow \citep{charnley90, begelman90, slavin93, pittard06}. We explore how this conclusion relates to the galactic wind as a whole in Section~\ref{sect:gal_wind}.

\citet{lord96} predict that the ionizing star clusters are well mixed with molecular matter in the form of clouds of characteristic sizes, $r_{\rm cl} \lesssim 1$~pc (Section~\ref{sect:intro}), thus providing copious sites from which gas can be evaporated and/or ablated. Thus, the fact that multiple PDR regions are predicted from far-IR spectroscopy, together with the observed close packing of the individual sources, the uniformity of the starburst conditions throughout the central regions, and the identification of a ubiquitous broad-line component, all point towards gas evaporation/ablation by cluster winds causing a highly turbulent velocity field being the root cause of the observed broad line widths.

%%%%%%%%%%%%%%%%%%%%%%%%%%%%%%%%%%
\section{Properties of the ionized gas} \label{sect:prop_gas}

\subsection{Flux Variations} \label{sect:results_flux}

\begin{figure*}
\centering
\plotone{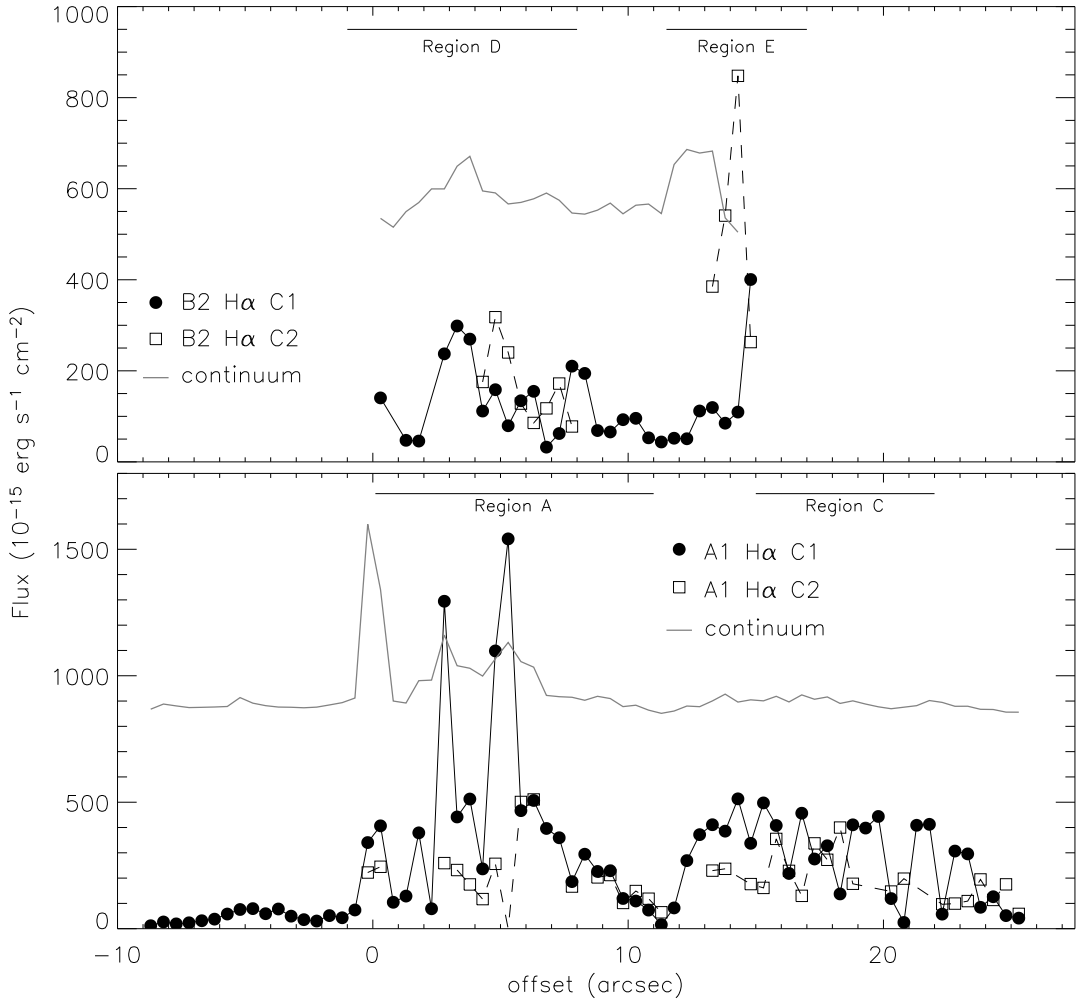}
\caption{H$\alpha$ flux for \textit{c1} and \textit{c2} measured from the A10 spectrum sets. This time we have split the measurements for the two slits up, with the slit A1 results in the lower panel and slit B2 results in the upper panel. The solid lines show the continuum level (in $10^{-15}$~erg$^{-1}$~s$^{-1}$~cm$^{-2}$, multiplied by 30 and offset by 800 for slit A1 and by 500 for slit B2) measured in the range 6620--6700~\AA. (A color version of this figure is available on-line.)}
\label{fig:ap10flux}
\end{figure*}

Fig.~\ref{fig:ap10flux} shows the distribution of the integrated H$\alpha$ line-flux in \textit{c1} and \textit{c2} along the section of the slits highlighted in white in Fig.~\ref{fig:finder}. Also plotted are the continuum flux levels (measured in the range 6620--6700~\AA) for each slit, multiplied and offset for ease of comparison (factors are listed in the figure caption). The relative continuum flux levels are consistent with the \textit{HST} F814W image, and clearly show the bright clumps A, D and E. These regions can also be seen in both the flux levels of \textit{c1} and \textit{c2}. In region A, the two peaks in \textit{c1} correspond well to peaks in the continuum brightness, but at the location of M82-A1 (offset = $0''$), the H$\alpha$ intensity does not peak so strongly. This could be due to a spatial sampling effect if the continuum and H$\alpha$ intensities are offset by less than one spatial resolution element, but if the levels shown are correct, it would suggest that the ionizing flux emerging from the core of region A is higher than at M82-A1, possibly indicating that the age of region A is even younger than that of M82-A1 \citepalias[6.5~Myr;][see also Section~\ref{sect:ews}]{smith06}.

The flux in \textit{c2} relative to \textit{c1} indicates the significance of the \textit{c2} emitting gas. \textit{c2} is less than or equal to half the intensity of \textit{c1} in the north-eastern half of region A and most of region C. \textit{c2} has an equal intensity to \textit{c1} in the south-western half of region A and in region D (but here the peak is spatially offset by $2''$). In region E, \textit{c2} peaks at twice the brightness of \textit{c1} (but again is spatially offset by $\sim$$2''$). A close examination of the line profiles at this position shows that \textit{c2} has likely been misidentified, and should really be assigned to \textit{c3}. However, the flux peaks are real, and are definitely offset from the continuum peak.

\subsection{Interstellar Extinction} \label{sect:extinc}

\begin{figure*}
\centering
\plotone{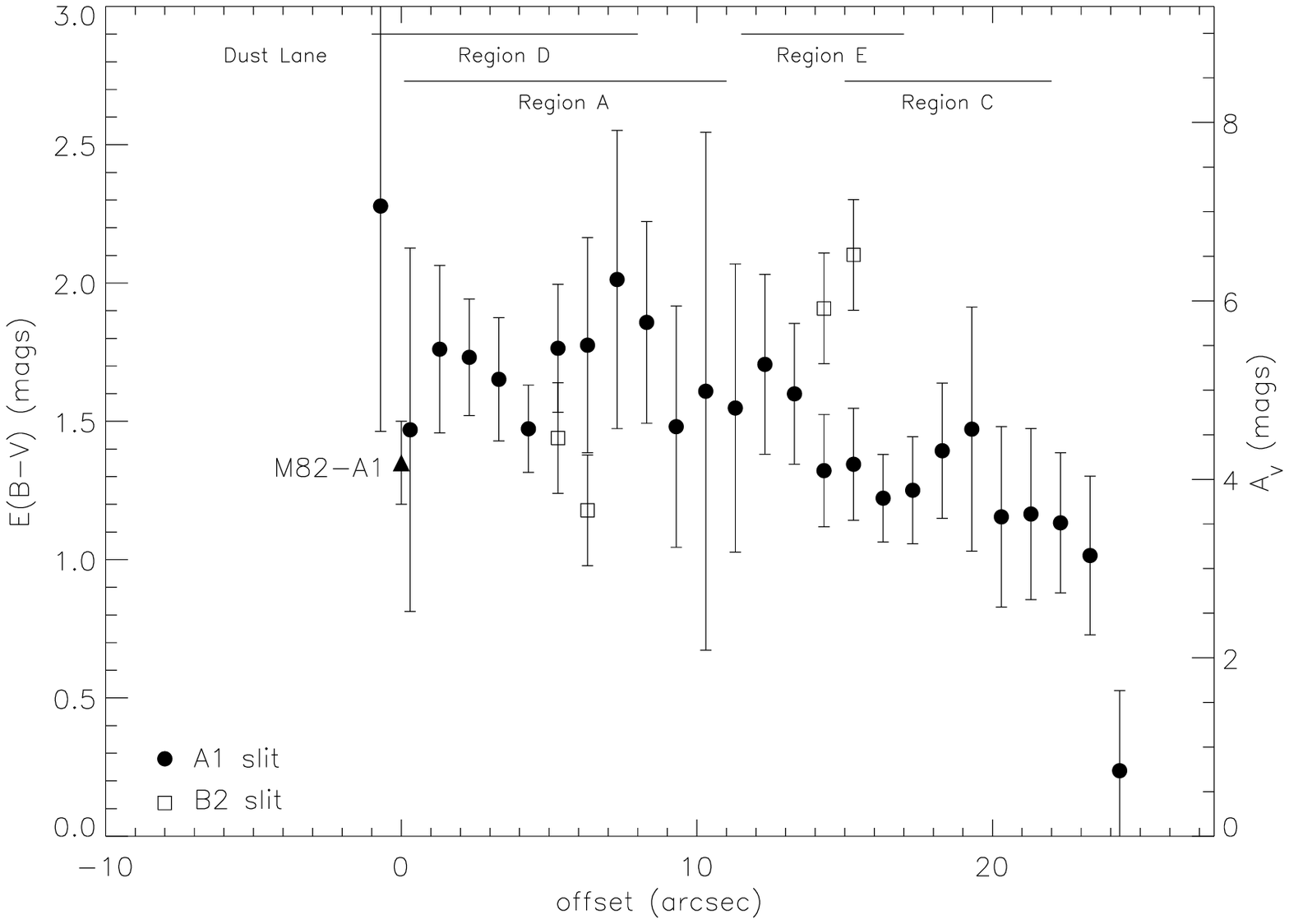}
\caption{Reddening values calculated from the H$\alpha$ to H$\beta$ line flux ratios, measured from the A20 spectra. The extinction is given in both $E(B-V)$ and $A_{V}$ units, and the $x$-axis scale is measured in arcseconds offset from the position of M82-A1 as per the previous figures. The triangle represents the reddening calculated for the H\two\ region surrounding M82-A1 \citepalias{smith06}.}
\label{fig:ap20ex}
\end{figure*}

The starburst core of M82 suffers from a high and non-uniform extinction. For example, several studies have found values for the visual extinction, $A_{V}$, of 2--12~mag from near-IR nebular line ratios (assuming a foreground screen model) \citep{satyapal95, forster01}. In this analysis we adopt a foreground screen model for simplicity, although we realise that this is likely to be unrealistic considering the level of dust present in the M82 disc \citep[values of $A_{V}\approx 20$--40~mag have been derived using mixed gas and dust models;][]{forster01}. One consolation is that at optical wavelengths, we can only see the least obscured part of the starburst, thus the effect of mixed dust will be fairly minimal. Using slits oriented along the minor-axis, \citet{heckman90} also measure the extinction in the central regions of M82, and find an excess to the north of the nucleus up to a radius of $\sim$300~pc, and that in the south, the extinction falls rapidly to a small and constant value beyond $R\approx 300$~pc. These findings are consistent with the assumed inclination of the galaxy.

To calculate the reddening distribution of the gas we compared the observed flux ratio of H$\beta$/H$\alpha$ (assuming no underlying stellar absorption) to the theoretical case B ratio given by \citet{humstor87}, using the Galactic extinction law of \citet{howarth83}. Shown in Fig.~\ref{fig:ap20ex} is the extinction in magnitudes measured at 20 pixel ($1''$) intervals along the slits. The uncertainty in the measurement of the H$\beta$ line flux dominates the error bars shown, and indeed only four extracted spectra from slit B2 had detectable H$\beta$ emission. For the few extracted spectra from slit A1 where we detect H$\gamma$, we also derived $A_{V}$ from the H$\gamma$/H$\beta$ ratio. The results were consistent with the extinction values obtained using the H$\alpha$/H$\beta$ ratio, although the errors were large.

The average extinction in region A is $A_{V}=5.5$~mag, which falls to an average of 4~mag in region C, and the data are consistent with a further decrease past region C towards the south-west. The extinction towards M82-A1 \citepalias{smith06} is less than that towards the majority of region A, perhaps explaining why this cluster stands out so clearly in optical images. We can only measure the reddening for one point to the north-east of M82-A1, and although the errors are large, it is consistent with being at a much higher value than the surroundings. This coincides with the beginning of a region of complete obscuration at optical wavelengths, extending towards the north-east of the starburst core (the dust-lane). Extinction measurements of the diffuse gas in this region ($-10$ to $-2$ arcsecs) are therefore made impossible due to the very low S/N or absence of the H$\beta$ lines. 

The extinction in slit B2 could only be measured for four points representing the brightest emission in the cores of regions D and E. In region D we find $A_{V}\approx 4$~mag, and is lower than that in region A. In region E, however, the extinction appears to be much higher at $A_{V}\approx 6$--6.5~mag.

\subsection{Electron Density}\label{sect:dens}

\begin{figure*}
\centering
\plotone{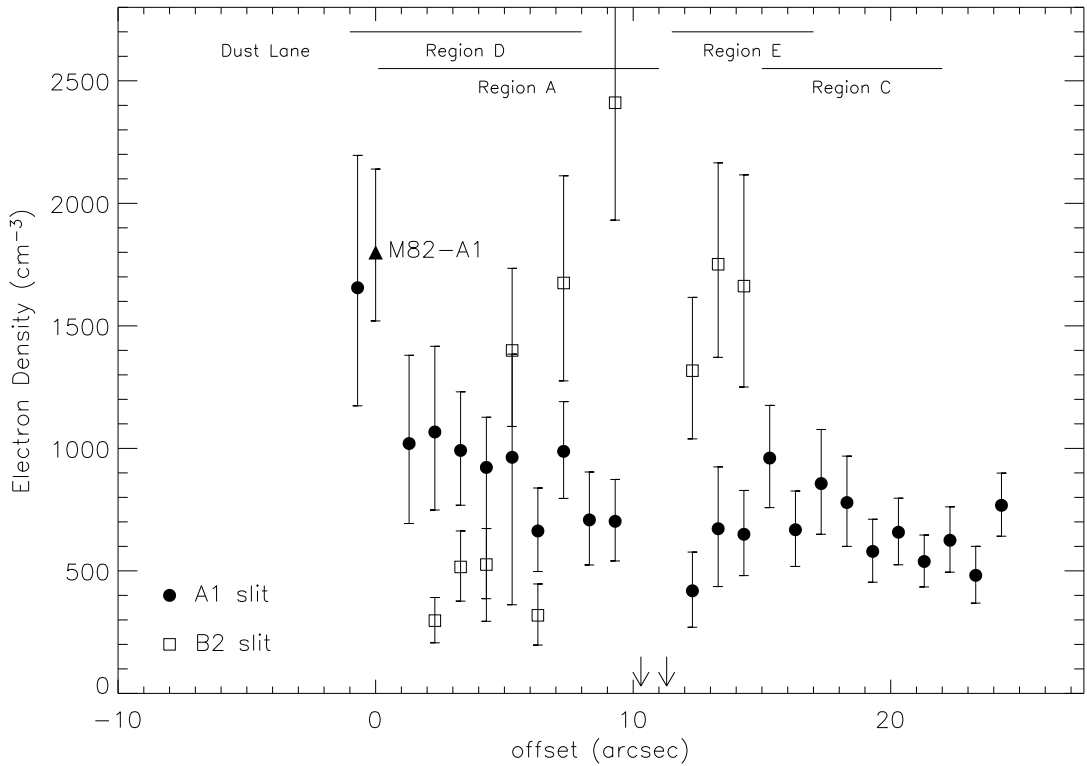}
\caption{Electron density derived from the [S\two]$\lambda 6717$/$\lambda 6731$ flux ratio, measured from the A20 spectrum sets. The triangle represents the electron density of the H\two\ region surrounding M82-A1 \citepalias{smith06}, and the arrows represent slit A1 measurements which fell below the low density limit for the [S\two] method.}
\label{fig:ap20dens}
\end{figure*}

We have measured the [S\two]$\lambda 6717$/$\lambda 6731$ line ratio by fitting Gaussian profiles to the emission lines in each of the A20 spectra for both slits. The distribution of derived electron densities, \Ne, (assuming an electron temperature, \Te{} = $10^{4}$~K) is plotted in Fig.~\ref{fig:ap20dens}.

In slit A1, we find the electron density to be highest around M82-A1 \citepalias[$n_{\rm e}=1800^{+340}_{-280}$~\cmt;][]{smith06}. [S\two] emission is not detected at all in spectra extracted from the north-east of M82-A1 (negative offsets), so in an attempt to derive an approximate average density for this region, we extracted a 60 pixel wide spectrum. However even summing over this many pixels did not yield a [S\two] detection, and therefore no density could be derived. From M82-A1, the density falls rapidly to $\sim$1000~\cmt\ in the centre of region A and to 600--800~cm$^{-3}$ in region C. Our observations provide constraints on the steepness of the density gradient from region A to A1: an enhancement of \Ne{} $\sim$ 1000~\cmt{} takes place over a projected distance of only 2.5~arcsecs (45~pc). In the area located between the two cluster complexes A and C, \Ne{} falls below the low-density limit ($<$100~\cmt) for the [S\two] ratio method, and is indicated by upper-limit arrows in Fig.~\ref{fig:ap20dens}. 

Our measurements are lower than those of \citet{oconnell78}, who also measured $n_{\rm e}$ from the [S\two] doublet ratio for regions A and C, and found an \emph{average} density of $n_{\rm e}=1800$~\cmt. The discrepancy may arise because at lower spatial-resolution, their data inevitably suffered from a higher degree of contamination and luminosity weighting than our high-resolution data, and were therefore biassed towards the densest gas. \citet{heckman90} also measured the electron density from their minor-axis slits, and found an \Ne{} of $\sim$1000~\cmt{} at the nucleus, falling to $\sim$100--300~\cmt{} at lateral distances of 1--2~kpc, which is in good agreement with our measurements.

We measure a very different distribution of densities in slit B2: close to the centre of region D\,\footnote{Note: slit B2 is not aligned with the major axis of region D and misses most of the eastern extent.}, the electron density is as low as 300~\cmt{}, but quickly rises to over 2000~\cmt{} in the area between regions D and E. Here slit B2 passes just to the north of a smaller cluster complex (2--4 arcsecs north of the 2.2~$\mu$m nucleus; Fig.~\ref{fig:finder}), which if really surrounded by gas at this density would be a very stifled region indeed. Alternatively, this density peak could result from scattered light from the starburst core/nucleus, under the assumption that higher pressures are associated with more intense zones of star formation. In region E, the electron density is also high with values of $\sim$1300--1700~\cmt, similar to that measured near M82-A1.

%%%%%%%%%%%%%%%%%%%%%%%%%%%%%%%%%%
\section{Physical state of the ionized ISM} \label{sect:state_ISM}

\subsection{Emission Measure} \label{sect:emission_meas}

In \citetalias{smith06} we discussed the surprising uniformity of the starburst over a large scale range by comparing our calculation of the ionization parameter of the M82-A1 H\two{} region to the results of the IR study by \citet{forster01}. We now use the observed H$\alpha$ fluxes to derive an estimate of the scale-length of the emitting regions to compare to and extend this discussion. 

A characteristic scale-length can be derived from the emission-measure, which is defined by:
\begin{equation}
F_{{\rm H}\alpha}^{\rm int} = \int n_{\rm e}^{2} \: \alpha_{\rm eff}^{{\rm H}\alpha} \: \frac{h \nu({\rm H}\alpha)}{4\pi D^{2}} \: \epsilon \: {\rm d}V,
\end{equation}
where $\alpha_{\rm eff}^{{\rm H}\alpha}$ is the case B recombination coefficient for H$\alpha$ \citep[$8.64\times 10^{-14}$~cm$^{3}$ s$^{-1}$;][]{osterbrock89}, $h \nu({\rm H}\alpha)$ is the energy of an H$\alpha$ photon, $D$ is the distance to the source, and $\epsilon$ is the gas volume filling factor. To remove the distance dependence, and to re-arrange in terms of a scale-length, d$V$ can be substituted with d$A$\,d$\ell$, where d$A = A / 4\pi D^{2}$ $\equiv$ angular area on the sky ($0.1\times 1$~arcsec in this case). Integrating over the whole emitting volume gives the mean intensity of H$\alpha$,
\begin{equation}
\big< \, F_{{\rm H}\alpha}^{\rm int} \, \big> = \big< \, n_{\rm e}^{2}\, \big> \: \alpha_{\rm eff}^{{\rm H}\alpha} \: \left<\epsilon\right> \: \ell \qquad \rm{erg~s^{-1}~cm^{-2}~arcsec^{-2}}.
\end{equation}
Hence, in reality what we can measure is $\ell(\epsilon=1)$ for all three H$\alpha$ line components in slit A1. Since $\epsilon \leq 1.0$, this is a lower-limit to the size-scales of the emitting H\two{} columns.

We find $\ell(\epsilon=1)$ to be less than 100~pc at all points, and is as low as $\sim$few parsecs to a few tens of parsecs in the north-east half of region A and the majority of region C. The consistency between line components shows that the gas from which each is emitted has similar characteristic sizes. Within the south-western half of region A and the A--C inter-clump region, the scale-lengths are larger, rising up to $\sim$50~pc.

The roughly constant values of the ionization parameter, log $U \approx -2.3$, within the main M82 starburst zone suggests that dust could be successfully competing with gas for ionizing photons 
on the surfaces of clouds \citep[see][]{dopita02}. In this case the \citeauthor{dopita02} model also suggests radiation pressure could be a significant factor. The luminosity densities in the M82 starburst zone, however, are well below those of the narrow line regions of AGN \citep[e.g.][]{cecil02a} or found in ULIRGs \citep[e.g.][]{thompson05} where models show that radiation acting on dust becomes important. Consistent with this view, we also find that the energy density in turbulent ISM motions exceeds the thermal pressure, and so even if radiation and thermal pressure are comparable, they are unlikely to be strongly influencing the M82 gas dynamics.

\subsection{Comparison to ISM models} \label{sect_disc_ism}

We can now discuss how our findings relate to the ISM models introduced in Section~\ref{sect:intro}, and what that implies for the M82 starburst. In the above section we found that the characteristic scale-size of the emitting regions remains below $\sim$50~pc ($3''$) throughout regions A and C. The uniformity of the starburst conditions must therefore stem from the fact that the scale of the emitting regions remains very small \citep[as suggested by][]{lord96}, hence the conditions are able to respond rapidly in order for the starburst to maintain such a near-constant state. However, within this overall uniformity there are variations. The smallest characteristic sizes ($\sim$few parsecs) are found in or near the dense cores of the clumps, whereas the largest sizes are found in the inter-clump region. These differences imply that the most compact clouds are found in the clump cores where the star-formation is most intense and the gas pressures are high.

\begin{figure*}
\centering
\begin{minipage}{7cm}
\includegraphics[width=7cm]{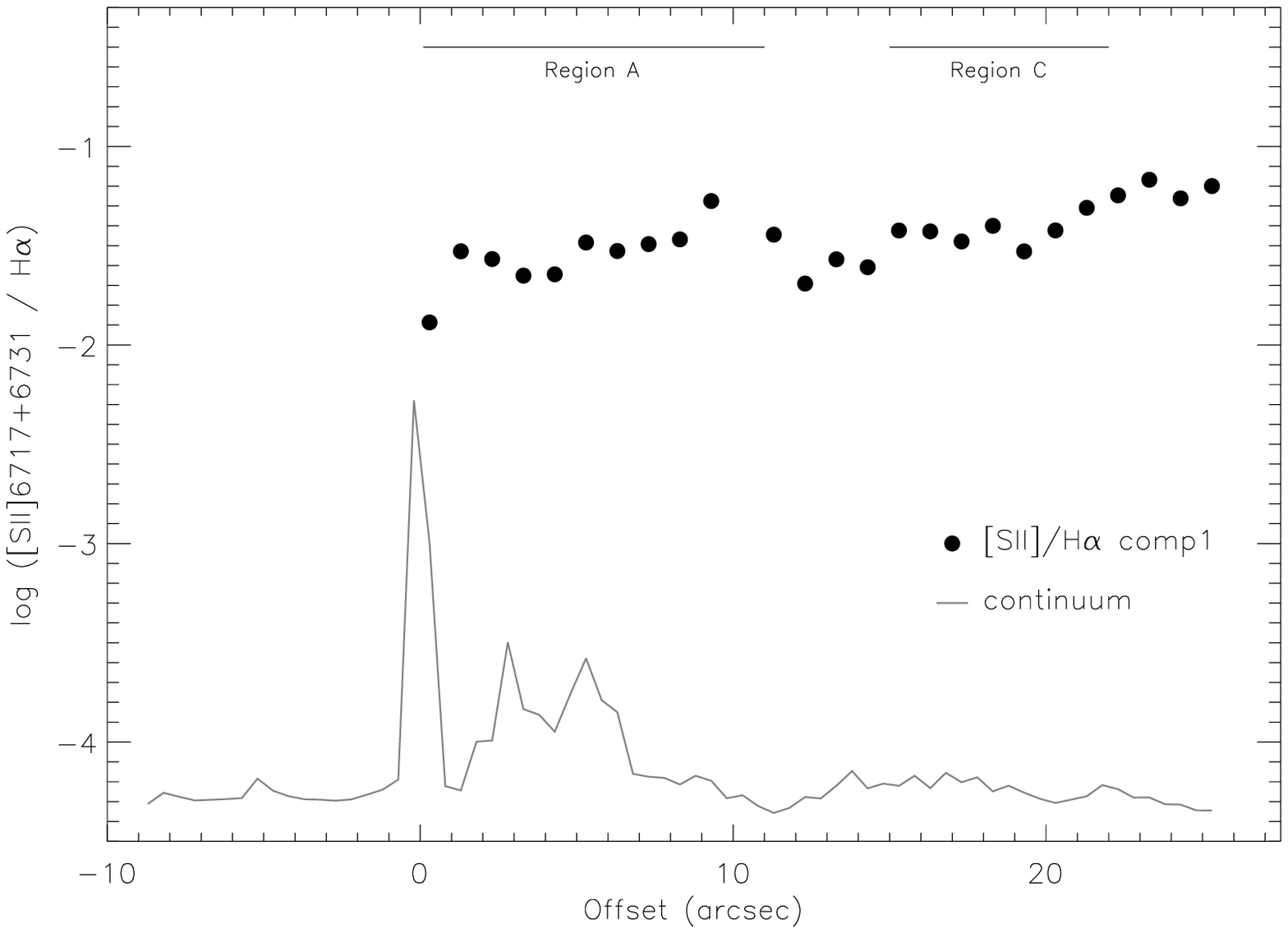}
\end{minipage}
\hspace*{0.3cm}
\begin{minipage}{7cm}
\includegraphics[width=7cm]{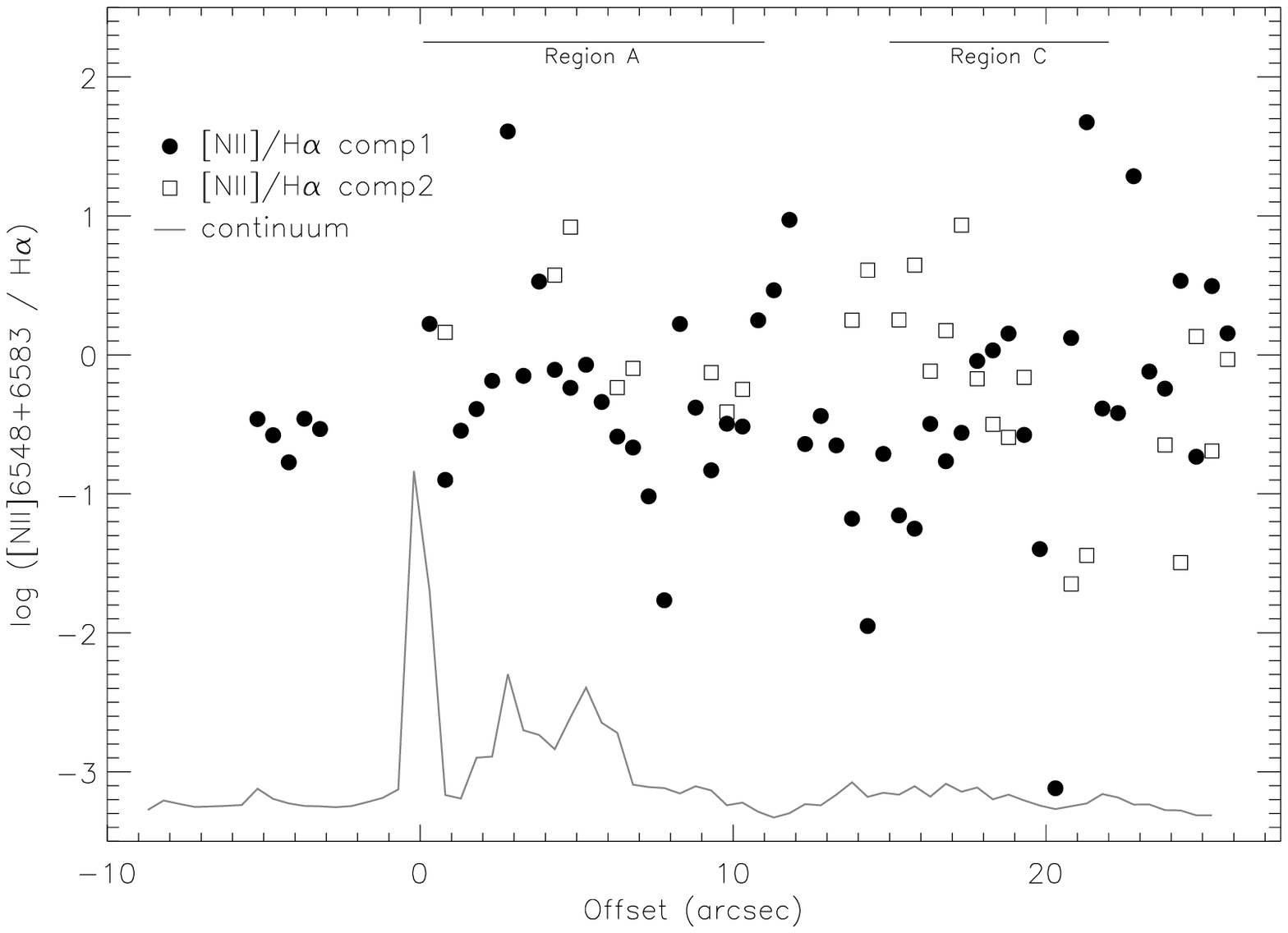}
\end{minipage}
\hspace*{0.3cm}
\caption{Plots showing the variation in the flux ratio of [S\two]$\lambda$6717+$\lambda$6731/H$\alpha$ (\emph{left}) and [N\two]$\lambda$6548+$\lambda$6583/H$\alpha$ (\emph{right}) with position on the A1 slit. The S/N of the H$\alpha$ and [NII] lines are high enough for the excitation of the two line components to be shown separately. For comparison, the continuum flux level is plotted as a solid line (arbitrary scale) on both plots.}
\label{fig:diagnostics}
\end{figure*}

Diagnostic flux ratios derived from the most commonly observed nebular lines can give a further indication of the ionization parameter and the mean level of ionization within the M82 starburst. The ratios of [S\two]$\lambda$6717+$\lambda$6731/H$\alpha$ and [N\two]$\lambda$6548+$\lambda$6583/H$\alpha$ are primarily tracers of the ionization parameter of the gas \citep{dopita00, dopita06b}, and are shown in Fig.~\ref{fig:diagnostics} as a function of position along slit A1. We find that the [S\two]/H$\alpha$ ratio changes very slowly across regions A and C, from $\sim$$-1.8$~dex at M82-A1, to $\sim$$-1$~dex to the south-west of region C. For [N\two]/H$\alpha$, we were able to measure the ratio independently for both \textit{c1} and \textit{c2} in many places, but find no significant difference between the two (except for perhaps in the inter-clump region where \textit{c2} is consistently higher than \textit{c1} for seven consecutive points). Again, the ratio stays roughly constant across regions A and C at an average of $-0.2\pm 0.5$~dex. As a comparison, \citet{shopbell98} measure equally high [N\two]/H$\alpha$ ratios ($-0.5$ to $-0.2$ dex) in the inner collimated zone for their low velocity component. Having access to the bluer nebular lines of [O\three] and H$\beta$ means that we can use their ratio as an indicator of the mean temperature ($T_{\rm eff}$) of the ionizing sources \citep{veilleux87, dopita00}. Unfortunately though, our ability to measure this ratio, like that of measuring the extinction, is limited by the low S/N of the H$\beta$ line. Where this line is detected, we find the flux ratio to remain constant within the uncertainties at log([O\three]/H$\beta$)~$\approx$~$-0.5$ over both regions A and C.

\citet{rigby04} discuss the effectiveness of certain optical and IR nebular line ratios, including [O\three]/H$\beta$, at tracing the mean temperature of the ionization source by comparing the consistency of the indicators between themselves and to models. They suggest that finding ratios that imply a low mean $T_{\rm eff}$ in starburst galaxies \citep[e.g.][this work]{thornley00} could result from either a lack of massive stars (M~$\gtrsim$~40~\Msun), or more likely, from the fact that the ultracompact H\two{} region phase is prolonged by the high densities and pressures in starburst ISMs. Under the influence of the high pressures, the most massive stars within the clusters may spend a significant proportion of their lifetimes embedded within the dense, highly-extincted \citep[$A_{V} \lesssim 50$~mag;][]{hanson02} regions in which they form, and they would remain undetectable, even through re-processed nebular optical or IR emission. Although their bolometric output may contribute to the surrounding gas for their entire main-sequence lifetime, it may influence nebular excitation indicators for a smaller fraction of this time \citep[$\sim$85 per cent;][]{rigby04}.

This situation agrees with what we know about the environment in M82, including its remarkably homogeneous ionization parameter \citep{forster01} and our detection of the dense, compact H\two\ region surrounding M82-A1 \citepalias{smith06}. Clearly not all clusters can be deeply embedded, since in many cases we can actually see the starlight from the cluster directly. However, the consistency in the line ratios (tracing nebular excitation and $T_{\rm eff}$) across the starburst regions must result from both the highly fractious nature of the ISM and the uniformity of the escaping radiation field. This radiation field must be attenuated by the high density gas to an extent to which it no longer reflects the presumably very inhomogeneous distribution of high-energy photons from the most massive stars within the clusters.

%%%%%%%%%%%%%%%%%%%%%%%%%%%%%%%%%%%%
\section{Structure of the M82 starburst} \label{sect:SB_struct}

\subsection{Evidence of the bar} \label{sect:bar}
We have used the high spectral- and spatial-resolution of our observations to track the kinematics of the H$\alpha$ line components across the two slits in unprecedented detail. We now use these measurements to provide constraints on the orbital parameters of the ionized gas, and the location of the individual starburst clumps within the galaxy.

\begin{figure}
\centering
\plotone{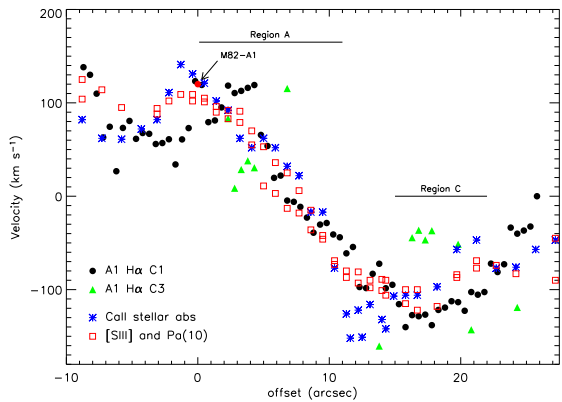}
\caption{H$\alpha$ \textit{c1} and \textit{c3} radial velocities for slit A1 plotted together with the near-IR stellar absorption line Ca\two\,$\lambda$8542 and the [S\three]$\lambda$9069 and Pa(10)$\lambda$9014 emission lines from \citet{mckeith93}. Offsets are measured relative to the position of M82-A1.}
\label{fig:ap10vel_mckeith}
\end{figure}

\begin{figure}
\centering
\plotone{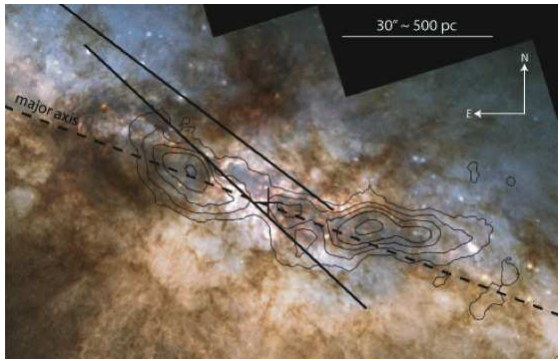}
\caption{Contours of integrated C$^{18}$O(J=1$\to$0) line intensity \citep{weis01} overlaid on an \textit{HST} broad-band colour-composite (blue: F439W, green: F555W, red: F814W; credit NASA/ESA R.\ de Grijs). The CO peaks trace the molecular/dust torus surrounding the stellar bar on either side of the nucleus. The location of the two STIS slits are marked with solid lines; the major axis is indicated with a dashed line and the cross marks the position of the 2.2~$\mu$m nucleus.}
\label{fig:degrijs_CO}
\end{figure}

As discussed in Section~\ref{sect:intro}, \citet{wills00} developed a model for the M82 bar system by analysing velocities derived from a number of neutral and ionized emission lines. They found the best-fitting bar model to have a peak radial velocity projected onto the sky of 140~\kms, an angular velocity of 217~\kms~arcsec$^{-1}$, a total length of 1~kpc, a core radius of 25~pc, but assume an opposite inclination to that which is commonly accepted (i.e.~$-80^{\circ}$). We now compare the results of their model to our observations to determine if we see evidence for any of the orbit families. In Fig.~\ref{fig:ap10vel_mckeith}, we plot the observed major-axis Ca\two{} stellar absorption and [S\three]$\lambda$9069 and Pa(10) nebular emission-line velocities from \citet{mckeith93} and our H$\alpha$ \textit{c1} and \textit{c3} velocities \citep[see also][]{greve02a}. The $x_{2}$-orbits, as traced by the stellar Ca\two{} measurements, can clearly be seen departing from the velocities determined from the gas emission-lines at offsets of $\sim$$+13''$ and $\sim$$0''$. Interestingly, M82-A1 appears to be located near the very top of the $x_{2}$-orbit pattern, unambiguously separated from the $x_{1}$-orbits. In other parts of the plot, the close match between the IR and optical data-sets indicates that the H$\alpha$-emitting gas follows the stellar rotation closely, although the rapidly varying extinction introduces inevitable complications. The inflection points at $\sim$0 and $\sim$+17$''$ represent the intersection between the $x_{2}$- and $x_{1}$-like orbits, where the gas is expected to shock and begin its journey towards the nucleus. Region C is clearly not associated with the $x_{2}$-like orbits, implying that it may be located within the molecular ring of material beyond the bar. 

\begin{figure*}
\centering
\plotone{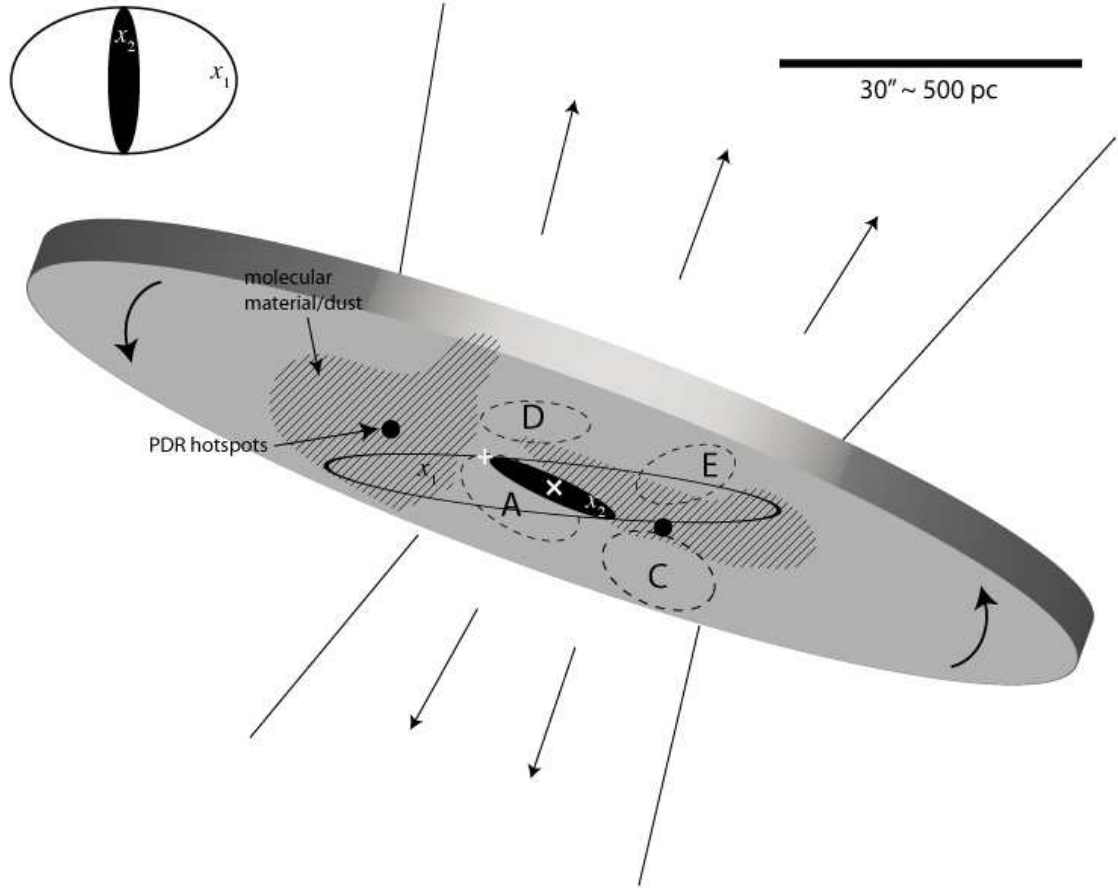}
\caption{Our proposed orientation of the M82 bar and disc, with respect to the main starburst clumps (labelled A, C, D and E) within the inner 2~kpc of the galaxy. The disc is inclined 80$^{\circ}$ to the line-of-sight and at a PA of 70$^{\circ}$. The bar $x_{1}$- and $x_{2}$-orbits are shown with an outlined and filled ellipse, respectively, and are also reproduced in plan-view in the upper-left. The major-axis of the bar extends along the $x_{1}$-orbits and is $\sim$1~kpc in length. At the ends of the bar, molecular material and dust (hatched regions) correspond to the location of the PDR hot-spots in the \citet{lord96} ISM model. The cluster M82-A1 (shown as a white plus-sign) is located at the end of an $x_{2}$-orbit with the maximum redshift observed. Clump C shows the highest blueshift indicating that it is located at the opposite end of the bar, whereas regions D and E must be located either behind or in front of the nucleus (shown as a white cross) due to the shallower velocity gradient observed between the two. The wind outflow cones are indicated above and below the disc. For comparison, it may be helpful to refer to Figs~\ref{fig:finder} and \ref{fig:degrijs_CO}.}
\label{fig:bar_orientation}
\end{figure*}

\subsection{Locating the starburst zones} \label{sect: }

Bearing in mind that the \citeauthor{wills00}~bar model assumed the opposite inclination to that which is generally accepted, we can now re-asses how the bar might be oriented with respect to the line-of-sight, and to the individual clumps. The \citeauthor{lord96}~ISM model predicts a concentration of molecular clouds (PDR regions) at either end of the stellar bar, which may be attributed to the interaction of bar $x_{1}$- and $x_{2}$-orbits causing the build-up of dust along the leading $x_{1}$-orbits \citep[a phenomenon sometimes termed `spraying';][]{athanassoula92b, wills00}. Fig.~\ref{fig:degrijs_CO} shows contours of integrated CO\,(J=1$\to$0) line intensity from \citet{weis01} overlaid on an \textit{HST} broad-band colour-composite, clearly showing how the peak of the CO emission corresponds to the location of the dust clouds either side of the nucleus and the proposed position of the two PDR hot-spots \citep{lord96}.

We can therefore associate this pile-up of molecular material/dust at either ends of the bar with the highly obscured regions to the north-east of clumps A and D and between clumps C and E extending out to the south-west. Our measurement of a rapid increase in extinction \citep[also found by][]{satyapal95, alonso03} and electron density to the north-east of regions A and D (see also Fig.~\ref{fig:degrijs_CO}), and the high extinction and density in region E and the D--E inter-clump region, is consistent with this conclusion. Evidence for a rapid fall-off in extinction to the south-west of region C may indicate the torus edge in this direction (Fig.~\ref{fig:ap20ex}). The size and location of these dust clouds is consistent with the predicted bar length ($\sim$1~kpc), and the comparative levels of extinction and prominence of the north-eastern dust cloud (again see Fig.~\ref{fig:degrijs_CO}) would imply that this is on the near-side of the starburst. This dust cloud also appears to wrap around region D and extend up into the northern inner-wind region -- whether or not this dust is being entrained into the wind flow at this point would require detailed kinematical observations.

In the region between clumps A and C, we do not measure a discontinuity in the extinction, but we do find a sudden drop in the electron density (Fig.~\ref{fig:ap20dens}). This suggests that there is a physical `hole' in the gas distribution between these two clumps and that the two regions are physically distinct, rather than the decrease in surface-brightness being due to an obscuring dust-lane. The remaining gas in this `void' must still be kinematically associated with the two clumps \citep[confirming the suggestion of][]{oconnell78} since there is no velocity discontinuity here. We can therefore now associate region C, located to the south of the bar region, as being outside the $x_{2}$-orbit region, and part of (or a result of star-formation within) the molecular torus surrounding the bar.

The fact that we can see evidence for both ends of the bar indicates that we are able to sample light from a considerable fraction of the M82 system. Contrary to previous models of the M82 system that assume large fractions of the M82 system are optically invisible \citep[e.g.][]{rieke93,satyapal97}, the main hidden region appears to be only in the middle of the bar near the nucleus. The most recent and perhaps most intense star-forming sites are located inside compact clouds spread throughout the system but concentrated in the torus surrounding the bar.

The fact that M82-A1 appears to be located at one end of the $x_{2}$-orbits places a strong constraint on the orientation of the bar within M82, and indicates that region A may have formed as a result of intense star-formation in the intersection between the $x_{1}$- and $x_{2}$-orbit families. The shallower velocity gradient between regions D and E indicate that they are located at a larger radius than that of regions A and C, but the distinction between these two regions is less clear than that between clumps A and C. Comparing Fig.~\ref{fig:finder} to Fig.~\ref{fig:degrijs_CO} shows that there is a great deal of extended emission and variable extinction in this inter-clump region, and the confused and variable velocities in this region indicate that we are seeing glimpses of star clusters through holes in a foreground dust screen. However, the fact that we find narrower line widths in regions D and E compared to A and C (Section~\ref{sect:results_fwhm}) supports the scenario of them being located at larger radii, since narrower line widths would indicate a more quiescent gas state, consistent with material further away from the nucleus.

Fig.~\ref{fig:bar_orientation} shows a schematic of our proposed spatial configuration of the starburst clumps and dust lanes in relation to the orientation of the bar orbits. This model implies that the eastern side of the bar is more distant, and that it extends from behind the dense dust cloud to the east of region A to just past region E. This orientation also implies that the eastern side of the inner $x_{2}$-orbits, where M82-A1 is located, is nearest.

\subsection{Implications for the production of the galactic wind} \label{sect:gal_wind}
In this section, we will use our observations together with the discussion given in the previous two sections to consider what implications our findings may have on how the superwind is produced in the starburst core. We will first consider how our spatially-resolved measurements of the gas density may affect the individual cluster winds, then go on to discuss possible explanations for the width of the observed line profiles, and finally how the dynamics of the broad component may indicate that we are seeing the roots of the large-scale galactic superwind.

\subsubsection{Densities and pressures}
In \citetalias{smith06}, we explored how the unusually high density of the gas surrounding M82-A1 has stifled the growth of its H\two{} region, and commented how it is hard to understand the development of the superwind if this cluster is representative of the entire starburst. We are now in a better position to know what the representative conditions are, and how these might vary over the face of the starburst. It is clear from high-resolution \textit{HST} images of the galactic wind (Gallagher et al., in prep.) that the outflow \emph{is} being driven directly from the central clumps (region A in particular), but our measured densities of the gas within these cluster complexes are still very high (500--900~\cmt, or $P/k \approx 0.5$--1.0$\times 10^{7}$~\cmt~K; \citetalias{smith06}) compared to standard Galactic ISM values. However, they are less than half of that found near M82-A1, and this highlights a possibly significant difference between a relatively isolated cluster such as M82-A1 and the densely-packed clusters in the clump cores. It appears that instead of having collimated winds from individual SSCs, the cluster complexes (clumps) produce a high pressure zone, or an 'energy injection centre', and the winds result from hot gas expanding out of these zones, upwards into the halo. This scenario will be explored in more detail by Gallagher et al. (in prep.).

\subsubsection{Dynamics of the broad line component}
We can now turn our attention to whether, by observing this broad, underlying component to the H$\alpha$ line, we are indirectly detecting the roots of the superwind flow. In our proposed hypothesis, \textit{c2} represents gas that has been stripped off the surface of gas clouds embedded in the starburst, so in principle the radial velocity offset between this component and the brighter (hence presumably denser) \textit{c1} gas can be used to trace the flow of the hot superwind into which it is being entrained.

Slit A1 passes through the very core of region A, and even though here we see some of the broadest line widths (up to 300~\kms), we do not see any velocity offset between the components, indicating that at this point there is still no preferred outflow direction. In region C, the slit passes through the southern side of this clump, rather than its core. Already at this radius, some evidence of an ordered wind flow is detected in the kinematical studies of \citet{mckeith95} and \citet{shopbell98}. These authors find that within 200~pc south of the nucleus the H$\alpha$ components are separated by $\sim$50~\kms. However, this is still within the chaotic zone of the outflow, and certainly well within the energy injection zone, where it is thought the wind flow is still subsonic but rapidly accelerating \citep{shopbell98}. In this region we find that \textit{c2} is redshifted with respect to \textit{c1} by $\sim$40~\kms{} and we also detect an $\sim$80~\kms{} redshifted \textit{c3} profile (see Fig.~\ref{fig:ap10vel}). These values are consistent with the aforementioned studies, and show that here we are beginning to see an ordered flow with a preferred direction, and that \textit{c2} is tracing the wind roots.

The region beyond clump C is interesting in a number of ways: here the \textit{c1} velocities rapidly decrease back to $v_{\rm sys}$, departing from the major-axis velocity measurements of \citet{mckeith93}, and the offset between \textit{c1} and \textit{c2} reverses direction, becoming blueshifted by $\sim$40~\kms{} relative to \textit{c1}. This far end of the slit is coincident with lateral ionized streamers pointing towards the south-west (see Fig.~\ref{fig:finder}), indicating that this region may be part of a distinct flow, separate to the main wind. If clump C is located within the molecular torus surrounding the bar, then this region may be less influenced by the bar's collimating effects, explaining the lateral flow.

%%%%%%%%%%%%%%%%%%%%%%%%%%%%%%%%%%%%%
\section{Summary} \label{sect:summary}

The use of high spatial resolution {\it HST} optical images and STIS spectra have enabled us to fill in a few more pieces of the large and complex jigsaw puzzle of M82. The data-set has allowed us to probe the ionized gas environment across the nuclear starburst in unprecedented detail. We have concentrated on mapping the ionized gas properties in the starburst core using two STIS slit pointings. One slit passes through the core of region A, and the southern edge of region C, whilst the other slit passes through part of region D and ends within region E.

\begin{itemize}
  \item By performing a comparison of our radial velocity measurements to previous observations and models, we confirm the presence of a stellar bar with an inner Lindblad resonance (ILR), resulting in a set of unique family of $x_{2}$-orbits perpendicular to the bar major-axis. The radial velocity of M82-A1 is consistent with being on an $x_{2}$-orbit at one end of the perpendicular ellipse, implying that region A may have formed as a result of intense star-formation at the intersection of the $x_{1}$- and $x_{2}$-orbits \citep[as predicted by models;][]{athanassoula92a}.
  \item Since the interaction of $x_{1}$-orbits with the perpendicular $x_{2}$-orbits is predicted to lead to a build-up of gas and dust along the leading $x_{1}$-orbits \citep{ab99}, we conclude that the prominent dust clouds to the north-east of regions A and D, and between regions C and E extending towards the south-west, must have been formed in this way. 
  \item By comparing CO observations to an \textit{HST} broad-band colour composite of M82, we identify that the peak of the molecular emission corresponds to the molecular torus surrounding the ionized ring \citep{achtermann95, weis01}, the proposed positions of the PDR hotspots from \citet{lord96}, and the prominent dust clouds located on either side of the nucleus. Since region C is clearly not associated with the $x_{2}$-orbits, we assume it to be located within this molecular torus.
  \item Our data are consistent with a physical distinction between clumps A and C, but gas between the clumps follows the same radial velocity pattern as the clumps, suggesting that all of this material is in the rotating disk of M82.
  \item A shallower velocity gradient between regions D and E leads us to believe that they are located at a larger radius to clumps A and C, supported by the finding of narrower line widths in these regions. However, the strong, patchy foreground extinction indicates that we may be seeing these cluster complexes through gaps in a thick foreground dust screen.
  \item By measuring the [S\two] doublet ratio along the slits, we find the gas densities (pressures) in the individual clumps to be high (500--900~\cmt; $P/k \approx 0.5$--$1.0\times 10^{7}$~\cmt~K), but significantly lower than that measured for the gas immediately surrounding M82-A1. We find that instead of seeing collimated winds from individual SSCs, the cluster complexes (clumps) each appear to produce a high pressure zone, or an 'energy injection centre', and that the winds result from hot gas expanding out of these zones. We will explore this scenario in more detail in a forthcoming contribution (Gallagher et al. in prep.).
  \item We extend the uniformity of the ISM conditions found by \citet{forster01} down to parsec-scales by deriving the emitting scale-sizes and by measuring the ionization parameter sensitive line-ratios of [S\two]/H$\alpha$ and [N\two]/H$\alpha$. We find the size-scale of the emitting regions to be always less than 50~pc along slit A1, and that the smallest regions (few pc) are in or near the dense cores of clumps A and C, implying that the most compact clouds are found where the star-formation is most intense. Our results confirm the prediction of uniform conditions throughout the starburst resulting from the highly fragmented nature of the ISM and/or the importance of dust in absorbing a significant fraction of the Lyman continuum luminosity. Our data also support the well-mixed nature of the embedded ionizing sources and ionized gas.
  \item Our line-profile analysis has shown that the widths of all but a few line components are $>$30~\kms. We have explored the possible mechanisms for producing the observed broad-line widths in both the narrow component (\textit{c1}) and the broad component (\textit{c2}). The most likely explanation for broadening of \textit{c1} is a combination of gravitationally induced virial effects and the stirring of the ISM through the intense star-formation activity. A base level of turbulent broadening ($\sim$30~\kms) exists over the whole starburst region, whilst the scatter to broader widths (up to 100~\kms) is likely to result from the presence of multiple unresolved kinematical components (e.g.~expanding shells) along the line-of-sight. The contribution from unresolved components is inevitable due to the presence of large numbers of densely packed SSCs.
  \item A ubiquitous broad component (150--250~\kms) is seen in all nebular lines with a high enough S/N. We argue that the broad line widths are produced by the interaction of high-energy, ionizing photons and fast-flowing winds from the star clusters with the cool gas clumps found throughout the starburst zone. As the wind impacts the surface of the clouds, evaporation and ablation of cloud material results in a highly turbulent velocity field which manifests as this broad component.
  \item In our model, the broad component represents turbulent gas stripped from the surface of gas clouds through the action of stellar winds and SN from the surrounding SSCs, so therefore could be used to track the hot superwind as the cooler material becomes entrained into the wind flow. We find no velocity offset between \textit{c1} and \textit{c2} in of region A, implying that here in the clump cores, the wind has yet to possess a preferred outflow direction. Where the slit passes through the southern edge of region C, we begin to see an ordered flow: \textit{c2} has a consistent redshifted offset, and a further redshifted narrow component (\textit{c3}) can be identified.
  \item Beyond clump C, the offset between the \textit{c1} and \textit{c2} velocities is reversed, and \textit{c1} departs from the radial velocity pattern measured along the major-axis. Since these measurements are coincident with a number of lateral south-west pointing streamers, this has lead us to believe that this region may not be associated with the main collimated flow. If clump C is located within the molecular torus surrounding the bar, then this region may be less influenced by the bar's collimating effects, explaining the lateral flow.
\end{itemize}

In conclusion, it seems a number of key ingredients are needed for the M82 starburst to exist. A central concentration of gas is required, together with clumps of dense star formation and high interstellar pressures. Of course, these three essential components do not exist in isolation, they link together and are to a larger or lesser extent the result of external influences. Large-scale galaxy encounters (which we know M82 has experienced) very often result in the formation of a bar, and a bar is a very efficient method to funnel gas into the nuclear regions in order to build up the required central gas concentration. Furthermore, both clumps and bars can be the products of disc instabilities that occur when discs are too dynamically cool i.e.~gas rich. We can therefore reason that the properties of the M82 starburst are all inextricably linked to one another. The study of one will always require a study (or at the least a good understanding) of the others.

Furthermore, we can now explain both the highly fragmented nature of the ISM and the near-uniform starburst conditions through the high densities and pressures within the starburst core, since high ambient densities result in compact, small, high density interstellar gas clouds. In high-pressured galaxies such as M82, the characteristic scale-sizes are therefore smaller than less-pressured systems, and this means that the gas conditions are able to respond quickly to any changes. This ability to respond rapidly results in the observed near-constant state of the starburst (seen though indicators such as ionization parameter and line diagnostics).

%%%%%%%%%%%%%%%%%%%%%%%%%%%%%%%
\section*{Acknowledgments}
We thank the referee, Mike Dopita, for his very insightful comments that led to some interesting discussion. We also thank David Stys and Linda Dressel at STScI for their help in reducing the STIS observations. The Image Reduction and Analysis Facility ({\sc iraf}) is distributed by the National Optical Astronomy Observatories which is operated by the Association of Universities for Research in Astronomy, Inc.\ under cooperative agreement with the National Science Foundation. {\sc stsdas} is the Space Telescope Science Data Analysis System; its tasks are complementary to those in {\sc iraf}. Support for program \#9117 was provided by NASA through a grant from the Space Telescope Science Institute, which is operated by the Association of Universities for Research in Astronomy, Inc., under NASA contract NAS 5-26555.

%%%%%%%%%%%%%%%%%%%%%%%%%%%%%%%%%%
% Bibliography
%%%%%%%%%%%%%%%%%%%%%%%%%%%%%%%%%%
\bibliographystyle{apj}
\bibliography{/Users/msw/Documents/work/Thesis/thesis/references}

\end{document}